\documentclass[a4paper,11pt]{article}
\usepackage[left=20mm,right=20mm,top=20mm,bottom=20mm]{geometry}
\usepackage{amsmath}
\usepackage{amsfonts}
\usepackage{amssymb}
\usepackage{amsbsy}
\usepackage{color,graphics}
\usepackage{setspace}
\usepackage{epsfig}
\usepackage{cite}
\usepackage{graphicx}
\usepackage{latexsym}
\usepackage{multicol}
\usepackage{footnote}
\renewcommand\textit[1]{{\color{blue}#1}}
\newtheorem{theorem}{Theorem}[section]

\newtheorem{example}[theorem]{Example}

\newtheorem{remark}[theorem]{Remark}

\numberwithin{equation}{section}
\begin{document}
\begin{center}
\LARGE{Hyperelastic bodies under homogeneous Cauchy stress induced by non-homogeneous finite deformations}\\
\vspace{0.3cm}
\large{L. Angela Mihai\footnote{Corresponding author: L. Angela Mihai, Senior Lecturer in Applied Mathematics, School of Mathematics, Cardiff University, Senghennydd Road, Cardiff, CF24 4AG, UK, Email: MihaiLA@cardiff.ac.uk}\quad and\quad Patrizio Neff\footnote{Patrizio Neff, Chair for Nonlinear Analysis and Modelling, Fakult\"{a}t f\"{u}r Mathematik, Universit\"{a}t Duisburg-Essen, Thea-Leymann Sra\ss e 9, 45141 Essen, Germany, Email: patrizio.neff@uni-due.de}}\\
\vspace{0.3cm}

October 5, 2016
\end{center}
\vspace{0.3cm}

\begin{abstract}
We discuss whether homogeneous Cauchy stress implies homogeneous strain in isotropic nonlinear elasticity. While for linear elasticity the positive answer is clear, we exhibit, through detailed calculations, an example with inhomogeneous continuous deformation but constant Cauchy stress. The example is derived from a non rank-one convex elastic energy. \\

\noindent \textbf{Mathematics Subject Classification:} 74B20, 74G65, 26B25.\\

\noindent \textbf{Keywords:} nonlinear elasticity; invertible stress-strain law; non-homogeneous deformations; rank-one connectivity; loss of ellipticity.
\end{abstract}

%%%%%%%%%%%%%%%%%%%%%%%%%%%%%%%%%%%%%%%%%%%%%%%%%%%%%%%%%%%%
%%%%%%%%%%%%%%%%%%%%   NEW SECTION  %%%%%%%%%%%%%%%%%%%%%%%%
%%%%%%%%%%%%%%%%%%%%%%%%%%%%%%%%%%%%%%%%%%%%%%%%%%%%%%%%%%%%
\section{Introduction}
In isotropic linear elasticity, it is plain to see that a homogeneous stress is always accompanied by
homogeneous strain, provided that the usual positive-definiteness assumptions on the elastic energy are required. Indeed, the linear elastic energy takes the form
\[
W_{\mathrm{lin}}(\nabla\textbf{u})=\mu\ \|\mathrm{dev}\ \mathrm{sym}\nabla\textbf{u}\|^2+\frac{\kappa}{2}\left[\mathrm{tr}\left(\mathrm{sym}\nabla\textbf{u}\right)\right]^2,
\]
where $\textbf{u}: B_{0} \to B$ is the displacement vector, $\boldsymbol{\varepsilon}=\mathrm{sym}\nabla\textbf{u}=\left[\nabla\textbf{u}+(\nabla\textbf{u})^{T}\right]/2$ is the infinitesimal strain tensor, $\mathrm{tr}(\boldsymbol{\varepsilon})=\varepsilon_{11}+\varepsilon_{22}+\varepsilon_{33}$ is the trace of the strain tensor, and
\[
\mathrm{dev}\ \boldsymbol{\varepsilon}=\boldsymbol{\varepsilon}-\frac{1}{3}\mathrm{tr}(\boldsymbol{\varepsilon})\textbf{I}
\]
is the deviatoric strain, with $\textbf{I}$ the tensor identity. In the above formulation, $\|\cdot\|$ denotes to Frobenius norm, hence, for a second order tensor $\textbf{A}$, this satisfies $\|\textbf{A}\|^2=\textbf{A}:\textbf{A}=\mathrm{tr}(\textbf{A}^T\textbf{A})$.

The corresponding stress-strain law is
\[
\boldsymbol{\sigma}=2\mu\ \mathrm{dev}\ \boldsymbol{\varepsilon}+\kappa\ \mathrm{tr}(\boldsymbol{\varepsilon})\ \textbf{I}.
\]
This relation is invertible if and only if the shear modulus satisfies $\mu> 0$, and similarly the bulk modulus satisfies $\kappa> 0$. We note that, if $\boldsymbol{\sigma}=\overline{\textbf{T}}$ is given, then $\boldsymbol{\varepsilon}=\mathrm{sym}\nabla\textbf{u}=\boldsymbol{\sigma}^{-1}(\overline{\textbf{T}})$ is uniquely determined, and moreover, if $\mathrm{sym}\nabla\textbf{u}=\mathrm{constant}=\boldsymbol{\sigma}^{-1}(\overline{\textbf{T}})\in\mathrm{Sym}(3)$, where $\mathrm{Sym}(3)$ is the set of symmetric matrices, then
\begin{eqnarray}
\nabla\textbf{u}(\textbf{X})&=&\boldsymbol{\sigma}^{-1}(\overline{\textbf{T}})+\textbf{A}(\textbf{X}),\qquad \textbf{A}(\textbf{X})\in\mathfrak{so}(3),\label{Eq:lingrad}
\end{eqnarray}
where $\mathfrak{so}(3)$ is the set of skew-symmetric matrices. This implies
\[
\underbrace{\mathrm{Curl}\ \nabla\textbf{u}}_{=0}=\underbrace{\mathrm{Curl}\ \boldsymbol{\sigma}^{-1}(\overline{\textbf{T}})}_{=0}+\mathrm{Curl}\ \textbf{A}(\textbf{x}),
\]
hence $\mathrm{Curl}\ \textbf{A}(\textbf{X})=0$, and therefore $\textbf{A}(\textbf{X})=\overline{\textbf{A}}=\mathrm{constant}$ \cite{Neff:2008:NM}. Altogether, we have that a constant stress tensor $\boldsymbol{\sigma}=\overline{\textbf{T}}$ implies the following representation for the displacement
\begin{eqnarray}
\textbf{u}(\textbf{X})&=&\left[\boldsymbol{\sigma}^{-1}(\overline{\textbf{T}})+\overline{\textbf{A}}\right]\textbf{X}+\overline{\textbf{b}},\label{Eq:linu}
\end{eqnarray}
where $\overline{\textbf{A}}\in \mathfrak{so}(3)$ is arbitrary and $\overline{\textbf{b}}\in\mathrm{R}^3$ is an arbitrary constant translation. Up to infinitesimal rigid body rotations and translations, the homogeneous displacement state is therefore uniquely defined through the constant stress field $\boldsymbol{\sigma}=\overline{\textbf{T}}$.

In nonlinear elasticity, the similar question of whether constant stress implies constant strain is
considerably more involved. One reason for this is the need to decide about the choice of the stress measure. Here, we focus on the ``true'' or Cauchy stress tensor. 

For a homogeneous isotropic hyperelastic body under finite strain deformation, the Cauchy stress tensor can be represented as follows \cite{Green:1970:GA,Green:1968:GZ,Ogden:1997,TruesdellNoll:2004}:
\begin{equation}\label{Eq:sigma}
\boldsymbol{\sigma}(\textbf{B})=\beta_{0}\ \textbf{I}+\beta_{1}\ \textbf{B}+\beta_{-1}\ \textbf{B}^{-1},
\end{equation}
where $\textbf{B}=\textbf{F}\textbf{F}^{T}$ is the left Cauchy-Green tensor, with the tensor $\textbf{F}=\nabla\varphi$ representing the deformation gradient, and the coefficients:
\begin{equation}\label{Eq:betas}
\beta_{0}=\frac{2}{\sqrt{I_{3}}}\left(I_{2}\frac{\partial W}{\partial I_{2}}+I_{3}\frac{\partial W}{\partial I_{3}}\right),\qquad
\beta_{1}=\frac{2}{\sqrt{I_{3}}}\frac{\partial W}{\partial I_{1}},\qquad
\beta_{-1}=-2\sqrt{I_{3}}\frac{\partial W}{\partial I_{2}}
\end{equation}
are scalar functions of the strain invariants:
\[
I_{1}(\textbf{B})=\mathrm{tr}\ \textbf{B},\qquad I_{2}(\textbf{B})=\frac{1}{2}\left[\left(\mathrm{tr}\ \textbf{B}\right)^{2}-\mathrm{tr}\ \textbf{B}^{2}\right]=\mathrm{tr}\left(\textrm{Cof}\ \textbf{B}\right),\qquad I_{3}(\textbf{B})=\det\textbf{B},
\]
with $W(I_{1}, I_{2}, I_{3})$ the strain energy density function describing the properties of the isotropic hyperelastic material.

If the material is incompressible, then the Cauchy stress takes the form:
\begin{equation}\label{Eq:sigma:inc}
\boldsymbol{\sigma}(\textbf{B})=-p\ \textbf{I}+\beta_{1}\ \textbf{B}+\beta_{-1}\ \textbf{B}^{-1},
\end{equation}
where $p$ is an arbitrary hydrostatic pressure.

The answer to whether constant Cauchy-stress implies constant Cauchy-Green tensor $\textbf{B} =\textbf{F}\textbf{F}^T$ would be easy to give if we could assume that the relation (\ref{Eq:sigma}) is invertible. That this relation may be invertible for a number of nonlinear elastic models (among them variants of Neo-Hookean or Mooney-Rivlin materials \cite{Ciarlet:1988,Ogden:1997}, and the exponentiated Hencky energy \cite{Ghiba:2015:GNS,Martin:2016:MGN,Neff:2015:NEM,Neff:2015:NGL,Neff:2015:NGLMS}) has recently been shown in \cite{Neff:2015:NGL}.

If invertibility holds in (\ref{Eq:sigma}), then we have a unique left Cauchy-Green tensor $\overline{\textbf{B}}\in\mathrm{Sym}^{+}(3)$ which satisfies
\begin{eqnarray}
\nabla\varphi\ (\nabla\varphi)^{T}&=&\overline{\textbf{B}}=\boldsymbol{\sigma}^{-1}(\overline{\textbf{T}}).
\end{eqnarray}
The latter implies (formally equivalent to the infinitesimal situation) that 
\begin{equation}
\varphi(\textbf{X})=\left(\overline{\textbf{V}}\ \overline{\textbf{R}}\right)\textbf{X}+\overline{\textbf{b}}=\left[\sqrt{\boldsymbol{\sigma}^{-1}(\overline{\textbf{T}})}\ \overline{\textbf{R}}\right]\textbf{X}+\overline{\textbf{b}},
\end{equation}
where $\overline{\textbf{R}}\in\mathrm{SO}(3)$ is an arbitrary constant rotation, $\overline{\textbf{b}}\in\mathbb{R}^{3}$ is an arbitrary constant translation, and $\overline{\textbf{V}}$ is the left principal stretch tensor satisfying $\overline{\textbf{V}}^2=\overline{\textbf{B}}$, and is uniquely determined from the given stress $\boldsymbol{\sigma}=\overline{\textbf{T}}$ \cite[p. 55]{Ciarlet:1988}.

While it is tempting to adopt invertibility of (\ref{Eq:sigma}) as a desirable feature of any ideal nonlinear elasticity law (at least for situations in which there is no loss stability), we refrain from imposing invertibility at present. Renouncing invertibility, in this paper, we consider the question if, and how, a homogeneous Cauchy stress tensor can be generated by non-homogeneous finite deformations. First, in Section~\ref{sec:geometry}, we provide an explicit and detailed construction of such situations on a specific geometry that allows for the deformation to be continuous and homogeneous in two different parts of the domain, connected by a straight interface, such that the two homogeneous deformations are rank-one connected. Then, in Section~\ref{sec:example}, we present an example of an isotropic strain energy function, such that, if a material is described by this function and occupies a domain similar to those analysed, then the expressions for the homogeneous Cauchy stress and the corresponding non-homogeneous strains can be written explicitly.

%%%%%%%%%%%%%%%%%%%%%%%%%%%%%%%%%%%%%%%%%%%%%%%%%%%%%%%%%%%%
%%%%%%%%%%%%%%%%%%%%   NEW SECTION  %%%%%%%%%%%%%%%%%%%%%%%%
%%%%%%%%%%%%%%%%%%%%%%%%%%%%%%%%%%%%%%%%%%%%%%%%%%%%%%%%%%%%
\section{Homogeneous stress induced by different deformations}\label{sec:geometry}

If the same Cauchy stress (\ref{Eq:sigma}) can be expressed equivalently in terms of two different homogeneous deformation tensors $\textbf{B}=\textbf{F}\textbf{F}^{T}$ and $\widehat{\textbf{B}}=\widehat{\textbf{F}}\widehat{\textbf{F}}^{T}$, such that $\textbf{F}\neq\widehat{\textbf{F}}$ and $\textbf{B}\neq\widehat{\textbf{B}}$, then the question arises whether it is possible for some part of the deformed body to be under the strain $\textbf{B}$ while another part is under the strain $\widehat{\textbf{B}}$. For geometric compatibility, we must assume that there exist two non-zero vectors $\textbf{a}$ and $\textbf{n}$, such that the Hadamard jump condition is satisfied as follows \cite{Ball:1987:BJ,Ball:1992:BJ}:
\begin{equation}\label{Eq:rank1an}
\widehat{\textbf{F}}-\textbf{F}=\textbf{a}\otimes\textbf{n},
\end{equation}
where $\textbf{n}$ is the normal vector to the interface between the two phases corresponding to the deformation gradients $\textbf{F}$ and $\widehat{\textbf{F}}$. In other words, $\textbf{F}$ and $\widehat{\textbf{F}}$ are rank-one connected, \emph{i.e.}
\begin{equation}\label{Eq:rank1}
\mathrm{rank}\left(\textbf{F}-\widehat{\textbf{F}}\right)=1.
\end{equation}

Here, we show that, under certain further restrictions, this type of non-homogeneous deformations leading to a homogeneous Cauchy stress are possible, and to demonstrate this, we uncover a class of such deformations by constructing them explicitly.

%%%%%%%%%%%%%%%%%%%%%%%%%%%%%%%%%%%%%%%%%%%%%%%%%%%%%%%%%%%%
\subsection{Elastostatic equilibrium}
\begin{figure}[htbp]
\begin{center}
\scalebox{0.25}{\includegraphics{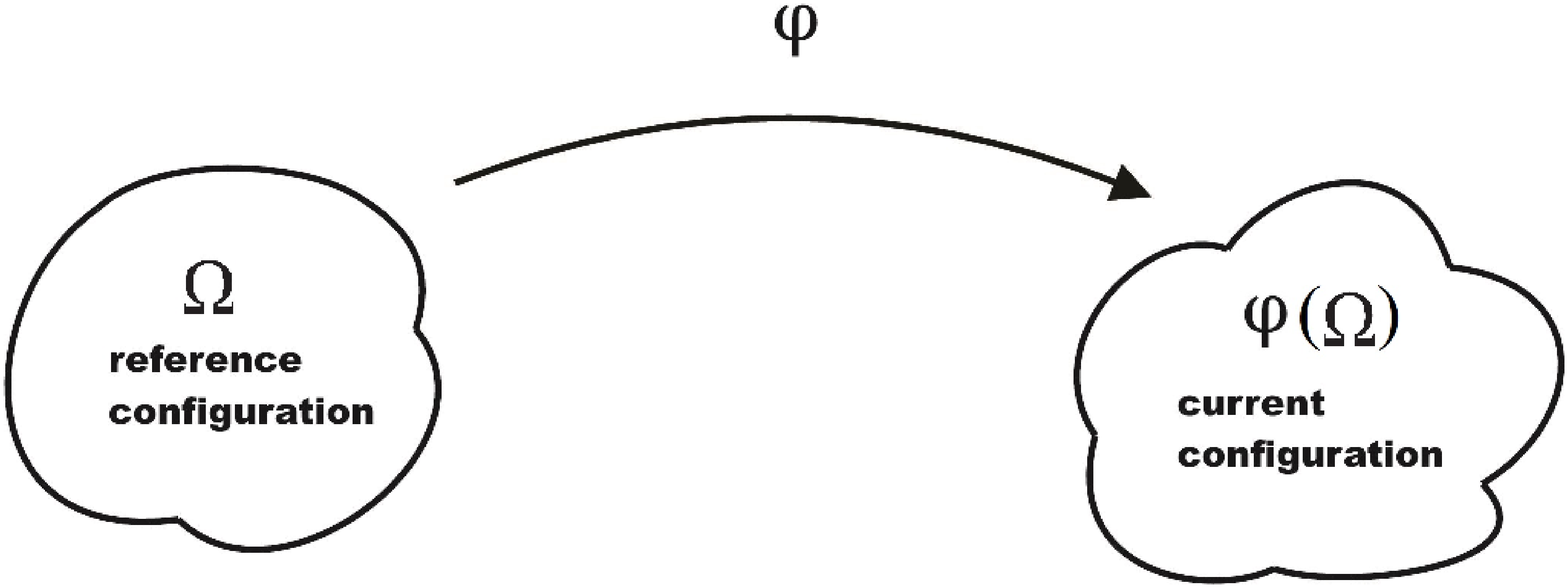}}
\caption{Schematic of elastic deformation.}\label{fig:deformation}
\end{center}
\end{figure}

A continuous material body occupies a compact domain $\bar\Omega$ of the three-dimensional Euclidean space $\mathbb{R}^{3}$, such that the interior of the body is an open, bounded, connected set $\Omega\subset\mathbb{R}^{3}$, and its boundary $\Gamma=\partial\Omega=\bar\Omega\setminus\Omega$ is Lipschitz continuous (in particular, we assume that a unit normal vector $\textbf{n}$ exists almost everywhere on $\Gamma$). The body is subject to a finite elastic deformation defined by the one-to-one, orientation preserving transformation
\[
\boldsymbol{\varphi}:\Omega\to\mathbb{R}^{3},
\]
such that  $J=\det\left(\nabla\boldsymbol{\varphi}\right)>0$ on $\Omega$  and $\boldsymbol{\varphi}$ is injective on $\Omega$ (see Figure~\ref{fig:deformation}). The injectivity condition on $\Omega$ guarantees that interpenetration of the matter is avoided. However, since self-contact is permitted, this transformation does not need to be injective on $\bar\Omega$.

%%%%%%%%%%%%%%%%
%%%%%%%%%%%%%%%

Let the spatial point $\textbf{x}=\boldsymbol{\varphi}(\textbf{X})$ correspond to the place occupied by the particle $\textbf{X}$ in the deformation $\boldsymbol{\varphi}$. For the deformed body, the equilibrium state in the presence of a dead load is described in terms of the Cauchy stress by the Eulerian field equation
\begin{eqnarray}
-\mathrm{div}\ \boldsymbol{\sigma}(\textbf{x})&=&\textbf{f}(\textbf{x}), \qquad \textbf{x}\in\varphi(\Omega).\label{Eq:equilibrium}
\end{eqnarray}
The above governing equation is completed by a constitutive law for $\boldsymbol{\sigma}$, depending on material properties, and supplemented by boundary conditions.

Since the domain occupied by the body after deformation is usually unknown, we rewrite the above equilibrium problem as an equivalent problem in the reference configuration where the independent variables are $\textbf{X}\in\Omega$. The corresponding Lagrangian equation of nonlinear elastostatics is
\begin{eqnarray}
-\mathrm{Div}\ \textbf{S}_{1}(\textbf{X})&=&\textbf{f}(\textbf{X}), \qquad \textbf{X}\in\Omega,\label{Eq:balance}
\end{eqnarray}
where $\textbf{S}_{1}=\boldsymbol{\sigma}\ \mathrm{Cof}\ \textbf{F}$ is the first Piola-Kirchhoff stress tensor, $\textbf{F}=\nabla\boldsymbol{\varphi}$ is the gradient of the deformation $\boldsymbol{\varphi}(\textbf{X})=\textbf{x}$, such that $J=\det\textbf{F}>0$, and $\textbf{f}(\textbf{X})=J\ \textbf{f}(\textbf{x})$.

For a homogeneous compressible hyperelastic material described by the strain energy function $W(\textbf{F})$, the first Piola-Kirchhoff stress tensor is equal to
\begin{eqnarray}
\textbf{S}_{1}(\textbf{F})=\frac{\partial W(\textbf{F})}{\partial\textbf{F}},\label{Eq:PK1}
\end{eqnarray}
and the associated Cauchy stress tensor takes the form $\boldsymbol{\sigma}=J^{-1}\textbf{S}_{1}\textbf{F}^{T}=\textbf{S}_{1}\left(\mathrm{Cof}\textbf{F}\right)^{-1}$.

The general boundary value problem (BVP) is to find the displacement $\textbf{u}(\textbf{X})=\varphi(\textbf{X})-\textbf{X}$, for all $\textbf{X}\in\Omega$, such that the equilibrium equation (\ref{Eq:balance}) is satisfied subject to the following conditions on the relatively disjoint, open subsets of the boundary $\{\Gamma_{D},\Gamma_{N}\}\subset\partial\Omega$, such that $\partial \Omega\setminus\left(\Gamma_{D}\cup\Gamma_{N}\right)$ has zero area \cite{LeTallec94,Mihai:2013:MG,Oden:1972}:
\begin{itemize}
\item On $\Gamma_{D}$, the Dirichlet (displacement) conditions
\begin{eqnarray}
\textbf{u}(\textbf{X})=\textbf{u}_{D}(\textbf{X}),\label{Eq:Dbc}
\end{eqnarray}
\item On $\Gamma_{N}$, the Neumann (traction) conditions
\begin{eqnarray}
\textbf{S}_{1}(\textbf{X})\textbf{N}=\textbf{g}_{N}(\textbf{X}),\label{Eq:Nbc}
\end{eqnarray}
where $\textbf{N}$ is the outward unit normal vector to $\Gamma_{N}$, and $\textbf{g}_{N}dA=\boldsymbol{\tau}da$, where $\boldsymbol{\tau}=\boldsymbol{\sigma}\textbf{n}$ is the surface traction measured per unit area of the deformed state.
\end{itemize}

The existence of a solution to the BVP depends on whether or not there exists a deformation which minimises, in the local or global sense, the total elastic energy of the body. Sufficient conditions that guarantee the existence of the global minimiser are that the strain energy density function is polyconvex, \emph{i.e.} convex as a function of deformation of line ($\textbf{F}$), of surface ($\mathrm{Cof}\ \textbf{F}$), and of volume ($\det\textbf{F}$) elements, and satisfies the coercivity (growth) and continuity requirements \cite{Ball:1977,Schroeder:2010:SN}. Clearly, in the absence of body forces, if the Cauchy stress is constant, then the equilibrium equation (\ref{Eq:equilibrium}) is satisfied.

%%%%%%%%%%%%%%%%%%%%%%%%%%%%%%%%%%%%%%%%%%%%%%%%%%%%%%%%%%%%
\subsection{Finite plane deformations}
\begin{figure}[htbp]
\begin{center}
\scalebox{0.42}{\includegraphics{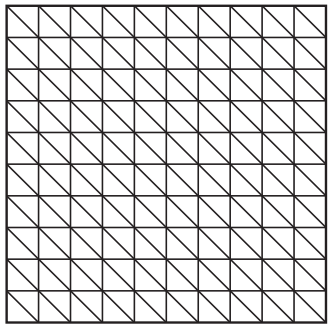}}
\caption{Uniform triangulation of elastic square.}\label{Fig:triangulation}
\end{center}
\end{figure}

Here, we consider the finite plane deformation of an elastic square partitioned into uniform right-angled triangles, as illustrated in Figure~\ref{Fig:triangulation}.

%%%%%%%%%%%%%%%%
%%%%%%%%%%%%%%%

Assuming that the deformation gradient is homogeneous on every triangle, in a single triangle $\Delta ABC$, the displacement field takes the general form
\begin{equation}\label{SP:eq:app:u}
\textbf{u}(\textbf{X})=\left[
\begin{array}{c}
u_{1}(\textbf{X})\\
u_{2}(\textbf{X})
\end{array}
\right]=
\left[
\begin{array}{c}
a_{11}X_{1}+a_{12}X_{2}+b_{1}\\
a_{21}X_{1}+a_{22}X_{2}+b_{2}
\end{array}
\right],
\end{equation}
where the six undetermined coefficients $a_{ij}$ and $b_{i}$, with $i,j=1,2$, are constants, and the associated deformation gradient is equal to
\begin{eqnarray}
\textbf{F}&=&\left[
\begin{array}{cc}
1+a_{11} & a_{12} \\
a_{21} & 1+a_{22}
\end{array}
\right].\label{Eq:F}
\end{eqnarray}

In order to determine the coefficients $a_{ij}$ and $b_{i}$, with $i,j=1,2$, we first evaluate the displacement (\ref{SP:eq:app:u}) at the three vertices $\{A, B, C\}\in\mathbb{R}^2$:
\[
A=\left[
\begin{array}{c}
X_{1}^A\\
X_{2}^A
\end{array}
\right],\qquad
B=\left[
\begin{array}{c}
X_{1}^B\\
X_{2}^B
\end{array}
\right],\qquad
C=\left[
\begin{array}{c}
X_{1}^C\\
X_{2}^C
\end{array}
\right].
\]
In this way, a system of six linear equations is obtained from which the six unknown coefficients are computed uniquely in terms of the  displacements $\left\{\textbf{u}^{A}, \textbf{u}^{B}, \textbf{u}^{C}\right\}\in\mathbb{R}^2$:
\[
\textbf{u}^A=\left[
\begin{array}{c}
u_{1}^A\\
u_{2}^A
\end{array}
\right],\qquad
\textbf{u}^B=\left[
\begin{array}{c}
u_{1}^B\\
u_{2}^B
\end{array}
\right],\qquad
\textbf{u}^C=\left[
\begin{array}{c}
u_{1}^C\\
u_{2}^C
\end{array}
\right],
\]
at the three vertices respectively, as follows:
\begin{eqnarray*}
a_{11}&=&\frac{u_{1}^{A}\left(X_{2}^{B}-X_{2}^{C}\right)+u_{1}^{B}\left(X_{2}^{C}-X_{2}^{A}\right)+u_{1}^{C}\left(X_{2}^{A}-X_{2}^{B}\right)}{X_{1}^{A}\left(X_{2}^{B}-X_{2}^{C}\right)+X_{1}^{B}\left(X_{2}^{C}-X_{2}^{A}\right)+X_{1}^{C}\left(X_{2}^{A}-X_{2}^{B}\right)},\\
a_{12}&=&\frac{u_{1}^{A}\left(X_{1}^{C}-X_{1}^{B}\right)+u_{1}^{B}\left(X_{1}^{A}-X_{1}^{C}\right)+u_{1}^{C}\left(X_{1}^{B}-X_{1}^{A}\right)}{X_{2}^{A}\left(X_{1}^{C}-X_{1}^{B}\right)+X_{2}^{B}\left(X_{1}^{A}-X_{1}^{C}\right)+X_{2}^{C}\left(X_{1}^{B}-X_{1}^{A}\right)},
\end{eqnarray*}
\begin{eqnarray*}
a_{21}&=&\frac{u_{2}^{A}\left(X_{2}^{B}-X_{2}^{C}\right)+u_{2}^{B}\left(X_{2}^{C}-X_{2}^{A}\right)+u_{2}^{C}\left(X_{2}^{A}-X_{2}^{B}\right)}{X_{1}^{A}\left(X_{2}^{B}-X_{2}^{C}\right)+X_{1}^{B}\left(X_{2}^{C}-X_{2}^{A}\right)+X_{1}^{C}\left(X_{2}^{A}-X_{2}^{B}\right)},\\
a_{22}&=&\frac{u_{2}^{A}\left(X_{1}^{C}-X_{1}^{B}\right)+u_{2}^{B}\left(X_{1}^{A}-X_{1}^{C}\right)+u_{2}^{C}\left(X_{1}^{B}-X_{1}^{A}\right)}{X_{2}^{A}\left(X_{1}^{C}-X_{1}^{B}\right)+X_{2}^{B}\left(X_{1}^{A}-X_{1}^{C}\right)+X_{2}^{C}\left(X_{1}^{B}-X_{1}^{A}\right)},
\end{eqnarray*}
\begin{eqnarray*}
b_{1}&=&\frac{u_{1}^{A}\left(X_{1}^{B}X_{2}^{C}-X_{1}^{C}X_{2}^{B}\right)+u_{1}^{B}\left(X_{1}^{C}X_{2}^{A}-X_{1}^{A}X_{2}^{C}\right)+u_{1}^{C}\left(X_{1}^{A}X_{2}^{B}-X_{1}^{B}X_{2}^{A}\right)}{X_{1}^{A}\left(X_{2}^{B}-X_{2}^{C}\right)+X_{1}^{B}\left(X_{2}^{C}-X_{2}^{A}\right)+X_{1}^{C}\left(X_{2}^{A}-X_{2}^{B}\right)},\\
b_{2}&=&\frac{u_{2}^{A}\left(X_{1}^{B}X_{2}^{C}-X_{1}^{C}X_{2}^{B}\right)+u_{2}^{B}\left(X_{1}^{C}X_{2}^{A}-X_{1}^{A}X_{2}^{C}\right)+u_{2}^{C}\left(X_{1}^{A}X_{2}^{B}-X_{1}^{B}X_{2}^{A}\right)}{X_{2}^{A}\left(X_{1}^{C}-X_{1}^{B}\right)+X_{2}^{B}\left(X_{1}^{A}-X_{1}^{C}\right)+X_{2}^{C}\left(X_{1}^{B}-X_{1}^{A}\right)}.
\end{eqnarray*}

For example, when $X_{1}^{A}=X_{1}^{C}=X_{1}^{B}-h=(j-1)h$ and $X_{2}^{A}=X_{2}^{B}=X_{2}^{C}-h=(i-1)h$, with $i,j=1,\cdots,N+1$ and $h>0$, the deformation gradient in the triangle $\Delta ABC$ can be expressed as follows
\begin{eqnarray*}
\textbf{F}&=&
\left[
\begin{array}{cc}
1+(u_{1}^{B}-u_{1}^{A})/h & (u_{1}^{C}-u_{1}^{A})/h \\
(u_{2}^{B}-u_{2}^{A})/h & 1+(u_{2}^{C}-u_{2}^{A})/h
\end{array}
\right].
\end{eqnarray*}

Similarly, in a triangle $\Delta A'BC$, the displacement field takes the form
\begin{equation}\label{SP:eq:app:uhat}
\widehat{\textbf{u}}(\textbf{X})=\left[
\begin{array}{c}
\widehat{u}_{1}(\textbf{X})\\
\widehat{u}_{2}(\textbf{X})
\end{array}
\right]=
\left[
\begin{array}{c}
\widehat{a}_{11}X_{1}+\widehat{a}_{12}X_{2}+\widehat{b}_{1}\\
\widehat{a}_{21}X_{1}+\widehat{a}_{22}X_{2}+\widehat{b}_{2}
\end{array}
\right],
\end{equation}
where the coefficients $\widehat{a}_{ij}$ and $\widehat{b}_{i}$, with $i,j=1,2$, are uniquely computed in terms of the displacements $\left\{\widehat{\textbf{u}}^{A'}, \widehat{\textbf{u}}^{B}, \widehat{\textbf{u}}^{C}\right\}\in\mathbb{R}^2$ at the three vertices $\{A', B, C\}\in\mathbb{R}^2$, respectively. Then the corresponding deformation gradient is equal to
\begin{eqnarray}
\widehat{\textbf{F}}&=&\left[
\begin{array}{cc}
1+\widehat{a}_{11} & \widehat{a}_{12} \\
\widehat{a}_{21} & 1+\widehat{a}_{22}
\end{array}
\right].\label{Eq:Fhat}
\end{eqnarray}

Given that the displacements are continuous at the vertices $B$ and $C$, \emph{i.e.} $\textbf{u}^{B}=\widehat{\textbf{u}}^{B}$ and $\textbf{u}^{C}=\widehat{\textbf{u}}^{C}$, the following two systems of algebraic equations are obtained:
\begin{eqnarray*}
a_{11}X_{1}^{B}+a_{12}X_{2}^{B}+b_{1}&=&\widehat{a}_{11}X_{1}^{B}+\widehat{a}_{12}X_{2}^{B}+\widehat{b}_{1},\\
a_{21}X_{1}^{B}+a_{22}X_{2}^{B}+b_{2}&=&\widehat{a}_{21}X_{1}^{B}+\widehat{a}_{22}X_{2}^{B}+\widehat{b}_{2},
\end{eqnarray*}
and
\begin{eqnarray*}
a_{11}X_{1}^{C}+a_{12}X_{2}^{C}+b_{1}&=&\widehat{a}_{11}X_{1}^{C}+\widehat{a}_{12}X_{2}^{C}+\widehat{b}_{1},\\
a_{21}X_{1}^{C}+a_{22}X_{2}^{C}+b_{2}&=&\widehat{a}_{21}X_{1}^{C}+\widehat{a}_{22}X_{2}^{C}+\widehat{b}_{2},
\end{eqnarray*}
from which the free coefficients $b_{i}$ and $\widehat{b}_{i}$, with $i=1,2$, can be written in terms of the coefficients $a_{ij}$ and $\widehat{a}_{ij}$, with $i,j=1,2$. Note that, though the displacements are continuous at the common vertices $B$ and $C$, the deformation gradient $\textbf{F}$ on the triangle $\Delta ABC$ may differ from the deformation gradient $\widehat{\textbf{F}}$ on the triangle $\Delta A'BC$.

In general, for a square partitioned into uniform right-angled triangles as depicted in Figure~\ref{Fig:triangulation}, given that the components of the displacement vector $\textbf{u}=[u_{1}, u_{2}]^{T}$ are continuous at every vertex $[X_{1}, X_{2}]^{T}$, for every interior vertex, there are $6$ local systems of algebraic equations of the form:
\begin{eqnarray*}
a_{11}X_{1}+a_{12}X_{2}+b_{1}&=&u_{1},\\
a_{21}X_{1}+a_{22}X_{2}+b_{2}&=&u_{2},
\end{eqnarray*}
one for each triangle meeting at that vertex, for every vertex on a side of the domain that is not a corner there are $3$ local systems, for two of the four corners there are $2$ systems, and for the other two corners there is only $1$ system of local equations.

Assuming that there are $2m^2$ right-angled triangles, we obtain $6(m-1)^2$ systems of two algebraic equations each corresponding to the interior vertices, $12(m-1)$ systems for the vertices situated on the boundaries that are not at the corners, $4$ systems for two of the four corners, and $2$ systems for the remaining two corners. This gives a total of $6(m-1)^2+12(m-1)+4+2=6m^2$ systems of two algebraic equation each, \emph{i.e.} $12m^2$ algebraic equations. For every triangle, there are $6$ unknown coefficients of the form $a_{ij}$ and $b_{i}$, with $i,j=1,2$, hence $12m^2$ such coefficients for the entire domain, which can be determined uniquely in terms of the displacements from the $12m^2$ algebraic equations.

It remains to find the equations from which the displacements are computed. Given the continuity of the displacement fields at the $(m+1)^2$ vertices, there are $2$ displacement components $u_{i}$, with $i=1,2$, for every vertex, \emph{i.e.} $2(m+1)^2$ displacement components in total. After the boundary conditions are imposed, $4m$ systems of two algebraic equations each, \emph{i.e.} $8m$ algebraic equations in total are provided at the vertices situated on the boundary. This leaves $2(m+1)^2-8m=2(m-1)^2$ displacement components, corresponding to the interior vertices, for which additional information is needed. This information may come, for example, from the condition that, on each triangle which does not have a vertex on the boundary, the determinant of the deformation gradient is equal to some given positive constant $d$, which is always valid for incompressible materials, where $d=1$, and which generates the required $2(m-1)^2$ equations. 

Hence, the displacement fields, which are continuous at the vertices, and the corresponding deformation gradients, which may differ from one triangle to another, can be uniquely determined from the boundary conditions and the constraint that the deformation is isochoric, for example. Moreover, though the displacements are continuous at each vertex, the deformation gradient, and hence the left Cauchy-Green tensor, may differ from one triangle to another.

Thus, any extra condition, such as the rank-one connectivity of the deformation gradients on two triangles having a common edge would mean additional constraints on the solution, and must be taken into account a priori, when selecting the boundary conditions. To see this, let $\boldsymbol{\sigma}$ be a homogeneous Cauchy stress tensor given by (\ref{Eq:sigma:inc}), such that it can be expressed equivalently in terms of two different homogeneous tensors $\textbf{B}=\textbf{F}\textbf{F}^{T}$ and $\widehat{\textbf{B}}=\widehat{\textbf{F}}\widehat{\textbf{F}}^{T}$, where $\textbf{F}$ and $\widehat{\textbf{F}}$ take the form (\ref{Eq:F}) and $(\ref{Eq:Fhat})$, respectively, and are rank-one connected.

We represent the respective first Piola-Kirchhoff stress tensors (\ref{Eq:PK1}) as follows:
\[
\textbf{S}_{1}=\boldsymbol{\sigma}\ \mathrm{Cof}\ \textbf{F}
=\left[
\begin{array}{cc}
S_{11} & S_{12}\\
S_{21} & S_{22}
\end{array}
\right],\qquad
\widehat{\textbf{S}}_{1}=\boldsymbol{\sigma}\ \mathrm{Cof}\ \widehat{\textbf{F}}
=\left[
\begin{array}{cc}
\widehat{S}_{11} & \widehat{S}_{12}\\
\widehat{S}_{21} & \widehat{S}_{22}
\end{array}
\right].
\]

We wish to demonstrate that it is possible for an elastic body occupying a square domain to deform such that the deformation gradient is equal to $\textbf{F}$ on some part of the body and to $\widehat{\textbf{F}}$ on another part.

First, we notice that, for the square domain partitioned into right-angled triangles as discussed above, if the deformation gradient is $\textbf{F}$ on one set of triangles and $\widehat{\textbf{F}}$ on the remaining set, then the common vertices between the two sets must lie on the same straight line. To show this, we assume that there are three common vertices $[X^{(k)}_{1},X^{(k)}_{2}]^{T}$, with $k=1,2,3$, which are not co-linear. At each vertex, the displacements are continuous, \emph{i.e.} the following identities hold:
\begin{eqnarray*}
a_{11}X^{(k)}_{1}+a_{12}X^{(k)}_{2}+b_{1}&=&\widehat{a}_{11}X^{(k)}_{1}+\widehat{a}_{12}X^{(k)}_{2}+\widehat{b}_{1},\\
a_{21}X^{(k)}_{1}+a_{22}X^{(k)}_{2}+b_{2}&=&\widehat{a}_{21}X^{(k)}_{1}+\widehat{a}_{22}X^{(k)}_{2}+\widehat{b}_{2},\qquad k=1,2,3.
\end{eqnarray*}
Equivalently, we obtain six linearly independent homogeneous equations of the form:
\begin{eqnarray*}
X^{(k)}_{1}\left(a_{11}-\widehat{a}_{11}\right)+X^{(k)}_{2}\left(a_{12}-\widehat{a}_{12}\right)+b_{1}-\widehat{b}_{1}&=&0,\\
X^{(k)}_{1}\left(a_{21}-\widehat{a}_{21}\right)+X^{(k)}_{2}\left(a_{22}-\widehat{a}_{22}\right)+b_{2}-\widehat{b}_{2}&=&0, \qquad k=1,2,3,
\end{eqnarray*}
from which we deduce that $a_{ij}=\widehat{a}_{ij}$ and $b_{i}=\widehat{b}_{i}$, with $i,j=1,2$. Hence $\textbf{F}=\widehat{\textbf{F}}$.

\begin{remark}\label{Remark}
We conclude that, if the displacement field is continuous everywhere, and the deformation gradient is $\textbf{F}$ on one set of triangles and $\widehat{\textbf{F}}$ on the remaining set, then a single straight line separates the two sets, \emph{i.e.} the two sets are situated at opposite corners. In particular, there are no layers of the domain where these sets can alternate.
\end{remark}

 Next, we present some examples.

%%%%%%%%%%%%%%%%%%%%%%%%%%%%%%%%%%%%%%%%%%
\subsubsection{Non-homogeneous deformation}

We consider an elastic material occupying the unit square $\Omega=(0,1)\times(0,1)$, and satisfying the equilibrium equation (\ref{Eq:balance}) with the boundary conditions defined as follows.
\begin{itemize}
\item Non-homogeneous Dirichlet boundary conditions:
\begin{eqnarray}
\textbf{u}(\textbf{X})=
\left[
\begin{array}{c}
a_{11}X_{1}+a_{12}X_{2}\\
a_{21}X_{1}+a_{22}X_{2}
\end{array}
\right] & \mbox{on} & (0,1)\times \{0\}\cup\{0\}\times(0,1),\label{Eq:dbc1}\\
\widehat{\textbf{u}}(\textbf{X})=
\left[
\begin{array}{c}
\widehat{a}_{11}X_{1}+\widehat{a}_{12}X_{2}+\widehat{b}_{1}\\
\widehat{a}_{21}X_{1}+\widehat{a}_{22}X_{2}+\widehat{b}_{2}
\end{array}
\right] & \mbox{on} & (0,1)\times \{1\}\cup\{1\}\times(0,1),\label{Eq:dbc2}
\end{eqnarray}
such that the possibility of rigid body deformations is eliminated by assuming that the lower left-hand corner is clamped, \emph{i.e.} $u_{1}=u_{2}=0$ at $\textbf{X}=(0,0)$ (hence $b_{1}=b_{2}=0$).
\end{itemize}

\begin{figure}[htbp]
\begin{center}
(a)\ \scalebox{0.42}{\includegraphics{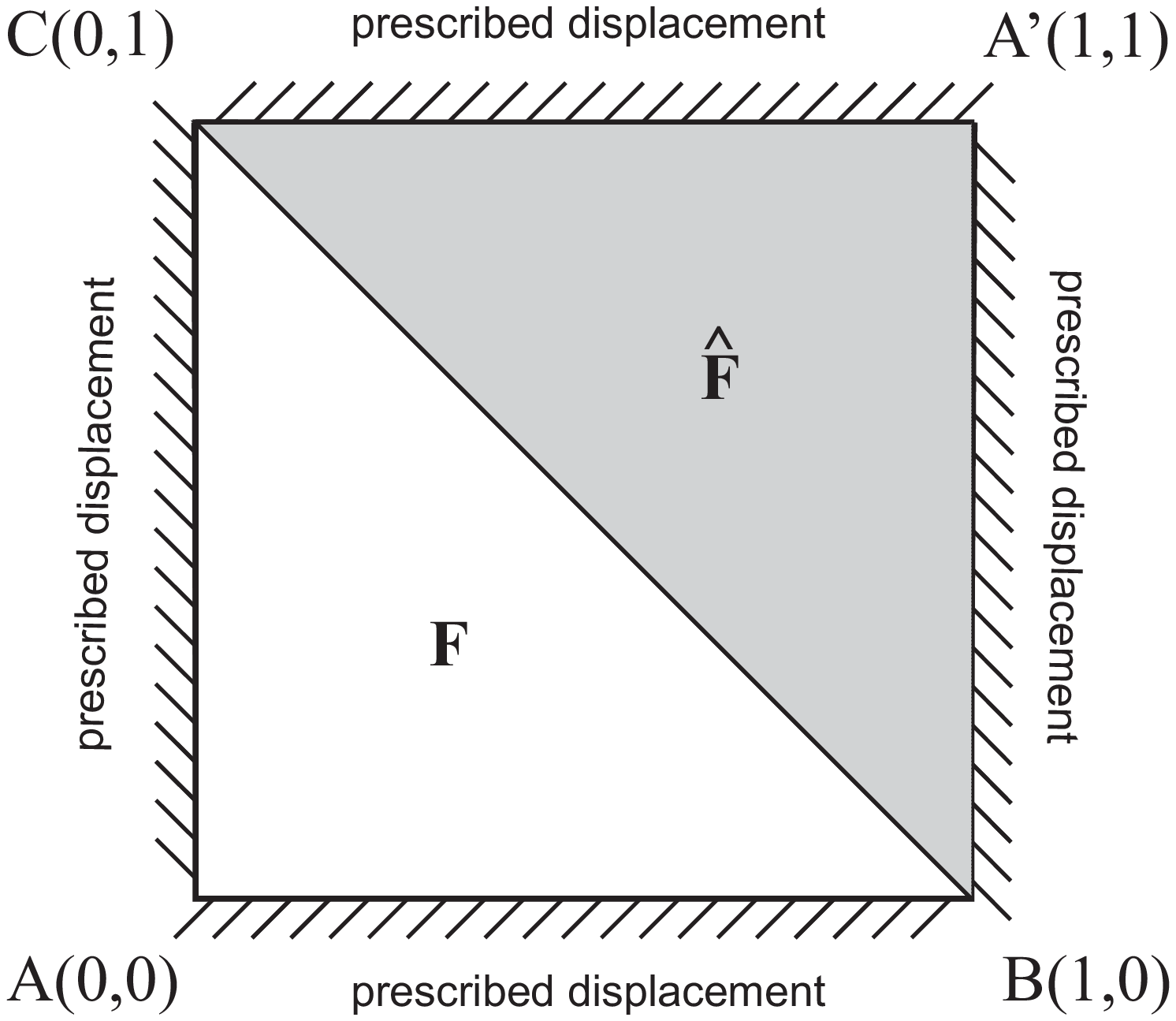}}\qquad
(b)\ \scalebox{0.42}{\includegraphics{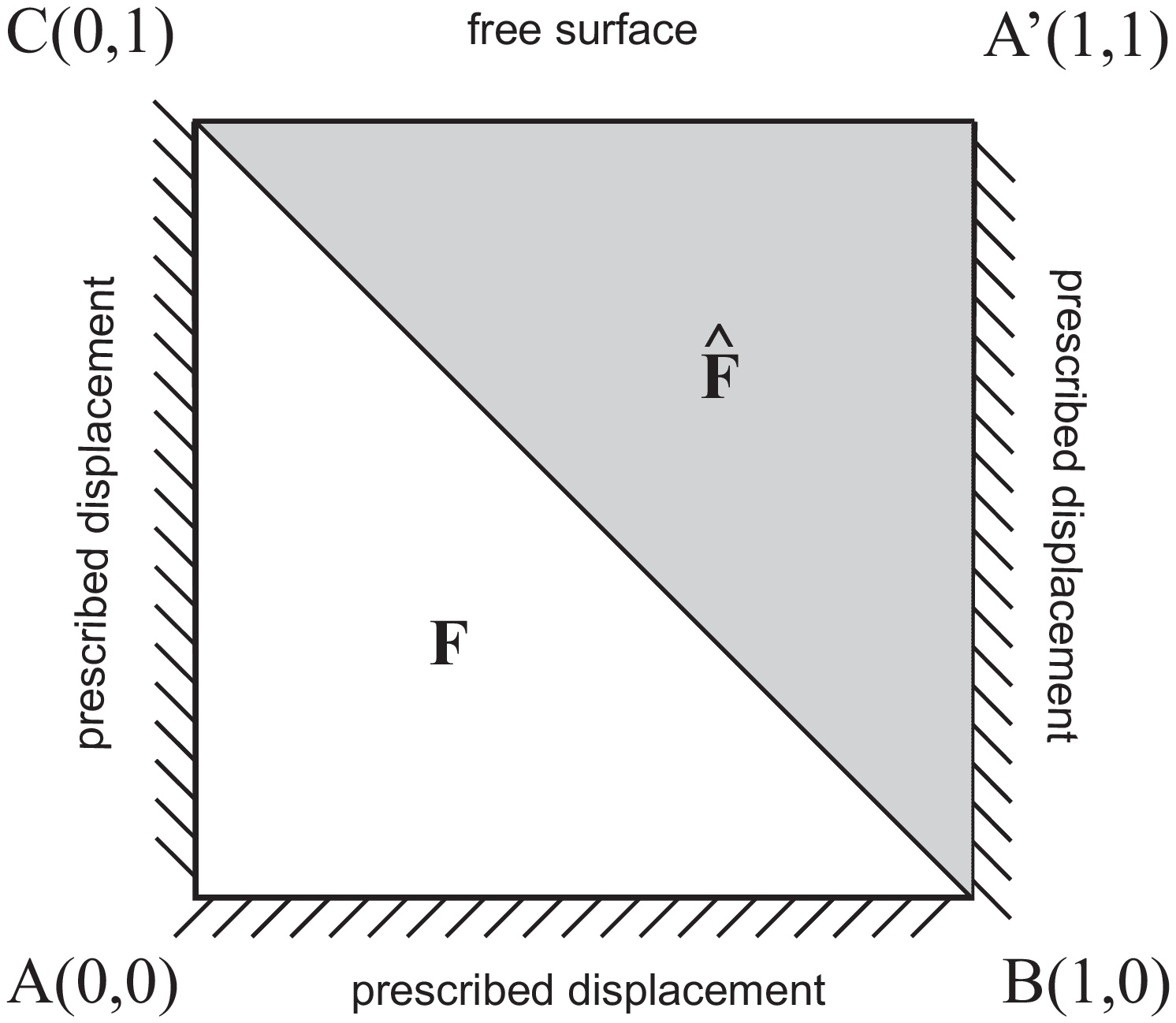}}
\caption{Elastic square with (a) Dirichlet boundary conditions and (b) with one side left free and Dirichlet boundary conditions on the other three sides, partitioned into two right-angled triangles with homogeneous deformation gradient in each triangle.}\label{fig:square-triangles-dbc}
\end{center}
\end{figure}

Solving the equation (\ref{Eq:balance}) with the Dirichlet boundary conditions (\ref{Eq:dbc1})-(\ref{Eq:dbc2}) yields the left Cauchy-Green tensor $\textbf{B}$ and the first Piola-Kirchhoff stress tensors $\textbf{S}_{1}$ in the triangular subdomain $\Delta ABC$, and the left Cauchy-Green tensor $\widehat{\textbf{B}}$ and the first Piola-Kirchhoff stress tensors $\widehat{\textbf{S}}_{1}$ in the subdomain $\Delta A'BC$, as illustrated in Figure~\ref{fig:square-triangles-dbc} (a). It follows that the given Cauchy stress tensor $\boldsymbol{\sigma}$ is the same throughout the deforming square.

%%%%%%%%%%%%%%%%
%%%%%%%%%%%%%%%%%

This non-homogeneous solution is also found when one side of the square is free and the Dirichlet boundary conditions (\ref{Eq:dbc1})-(\ref{Eq:dbc2}) are prescribed on the other three sides, as shown in Figure~\ref{fig:square-triangles-dbc} (b).

\begin{figure}[htbp]
\begin{center}
(a)\ \scalebox{0.42}{\includegraphics{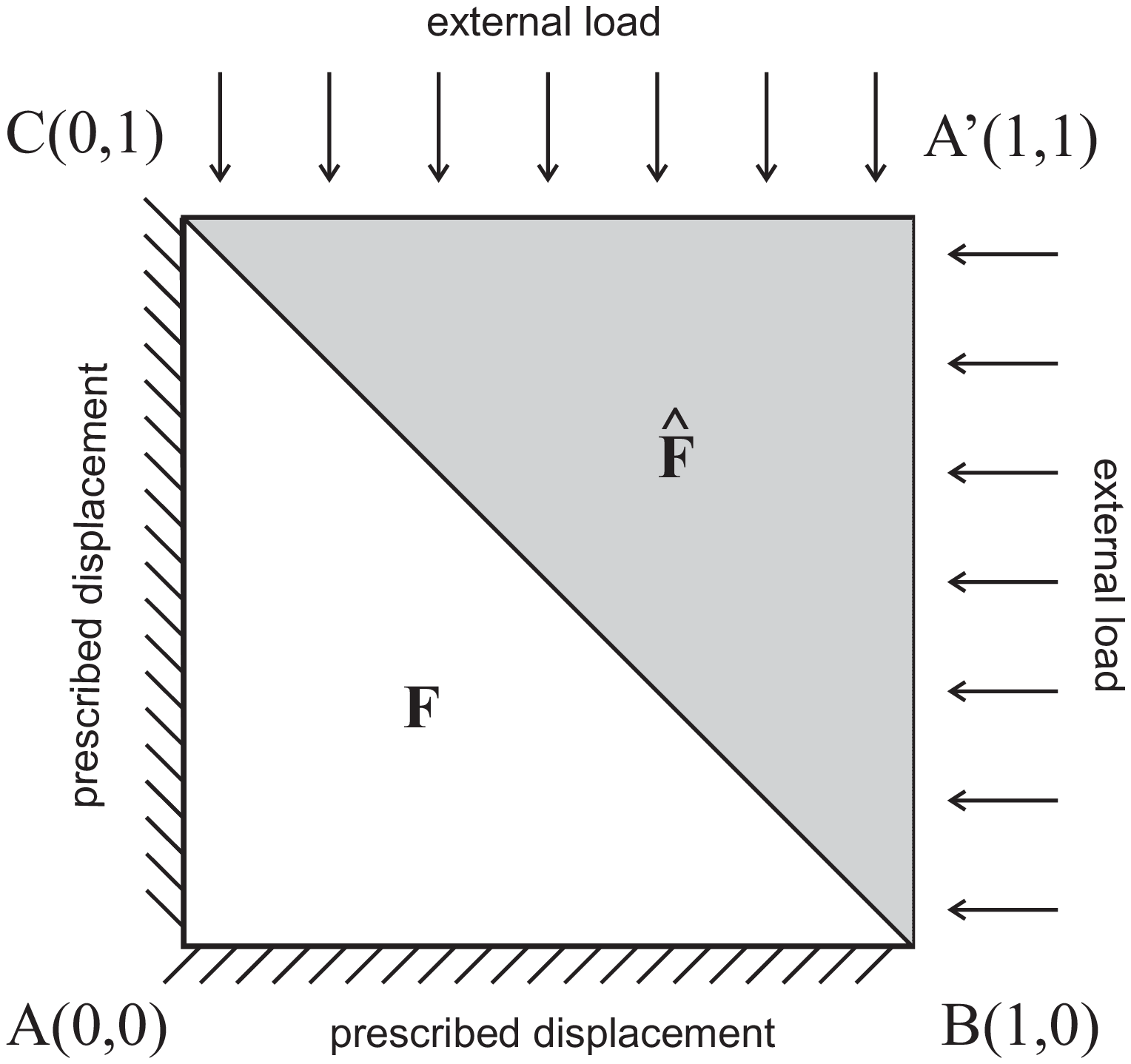}}\qquad
(b)\ \scalebox{0.42}{\includegraphics{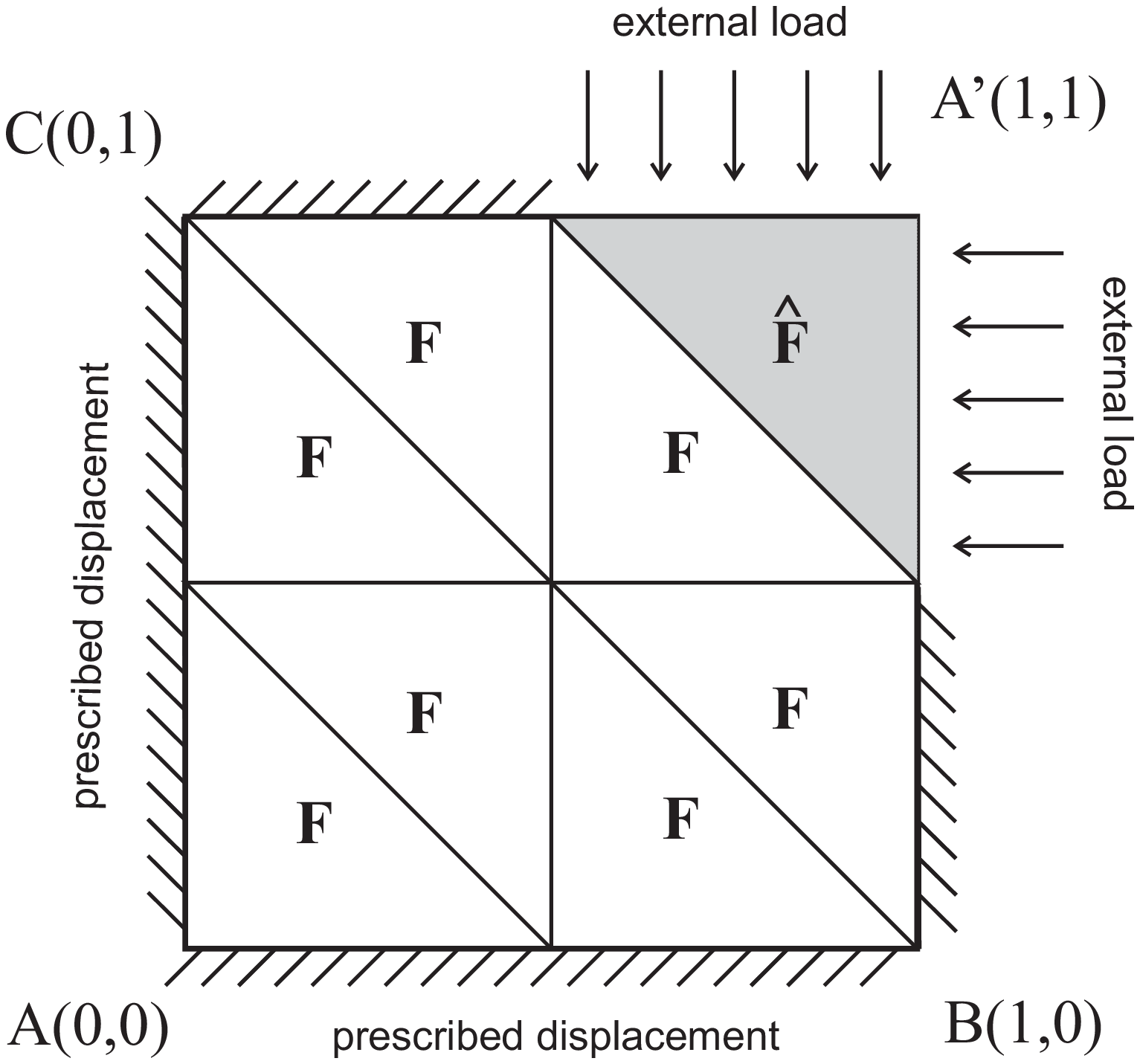}}
\caption{Elastic square with mixed boundary conditions, partitioned into right-angled triangles with homogeneous deformation gradient in each triangle.}\label{fig:square-triangles-mbc}
\end{center}
\end{figure}

\begin{itemize}
\item Alternatively, the above non-homogeneous solution can be attained under the following mixed boundary conditions, as indicated in Figure~\ref{fig:square-triangles-mbc} (a):
\begin{eqnarray}
\textbf{u}(\textbf{X})=
\left[
\begin{array}{c}
a_{11}X_{1}\\
a_{21}X_{1}
\end{array}
\right] & \mbox{on} & (0,1)\times \{0\},\label{Eq:mbc3:1}\\
\textbf{u}(\textbf{X})=
\left[
\begin{array}{c}
a_{12}X_{2}\\
a_{22}X_{2}
\end{array}
\right] & \mbox{on} & \{0\}\times(0,1),\label{Eq:mbc3:2}\\
\widehat{\textbf{S}}_{1}(\textbf{X})\left[
\begin{array}{r}
1\\
0
\end{array}
\right]=
\left[
\begin{array}{r}
\widehat{S}_{12}\\
\widehat{S}_{22}
\end{array}
\right]
& \mbox{on} & (0,1)\times \{1\},\label{Eq:mbc3:3}\\
\widehat{\textbf{S}}_{1}(\textbf{X})\left[
\begin{array}{r}
0\\
1
\end{array}
\right]=
\left[
\begin{array}{r}
\widehat{S}_{11}\\
\widehat{S}_{21}
\end{array}
\right] & \mbox{on} & \{1\}\times(0,1),\label{Eq:mbc3:4}
\end{eqnarray}
such that $u_{1}=u_{2}=0$ at $\textbf{X}=(0,0)$. Under these conditions, at a corner where one of the adjacent edges is subject to Dirichlet conditions and the other to Neumann conditions, the Dirichlet conditions take priority, and when both edges meeting at a corner are subject to Neumann conditions, these conditions are imposed simultaneously at the corner.
\end{itemize}

%%%%%%%%%%%%%%%%%
%%%%%%%%%%%%%%%

\begin{itemize}
\item However, a different non-homogeneous solution is obtained under the following mixed boundary conditions:
\begin{eqnarray}
\textbf{u}(\textbf{X})=
\left[
\begin{array}{c}
a_{11}X_{1}\\
a_{21}X_{1}
\end{array}
\right] & \mbox{on} & (0,1)\times \{0\},\label{Eq:mbc4:1}\\
\textbf{u}(\textbf{X})=
\left[
\begin{array}{c}
a_{12}X_{2}\\
a_{22}X_{2}
\end{array}
\right] & \mbox{on} & \{0\}\times(0,1),\label{Eq:mbc4:2}\\
\textbf{u}(\textbf{X})=
\left[
\begin{array}{c}
a_{11}X_{1}+a_{12}\\
a_{21}X_{1}+a_{22}
\end{array}
\right] & \mbox{on} & (0,1/2)\times \{1\},\label{Eq:mbc4:3}\\
\textbf{u}(\textbf{X})=
\left[
\begin{array}{c}
a_{11}+a_{12}X_{2}\\
a_{21}+a_{22}X_{2}
\end{array}
\right] & \mbox{on} & \{1\}\times(0,1/2),\label{Eq:mbc4:4}\\
\widehat{\textbf{S}}_{1}(\textbf{X})\left[
\begin{array}{r}
1\\
0
\end{array}
\right]=
\left[
\begin{array}{r}
\widehat{S}_{12}\\
\widehat{S}_{22}
\end{array}
\right]
& \mbox{on} & (1/2,1)\times \{1\},\label{Eq:mbc4:5}\\
\widehat{\textbf{S}}_{1}(\textbf{X})\left[
\begin{array}{r}
0\\
1
\end{array}
\right]=
\left[
\begin{array}{r}
\widehat{S}_{11}\\
\widehat{S}_{21}
\end{array}
\right] & \mbox{on} & \{1\}\times(1/2,1),\label{Eq:mbc4:6}
\end{eqnarray}
such that $u_{1}=u_{2}=0$ at $\textbf{X}=(0,0)$. As before, at a boundary point where one of the adjacent edges is subject to Dirichlet conditions and the other to Neumann conditions, the Dirichlet conditions take priority, and when both edges meeting at a point are subject to Neumann conditions, these conditions are imposed simultaneously at that point.
\end{itemize}

The solution of the equation (\ref{Eq:balance}) with the boundary conditions (\ref{Eq:mbc4:1})-(\ref{Eq:mbc4:6}) is illustrated schematically in Figure~\ref{fig:square-triangles-mbc} (b). Thus the given Cauchy stress $\boldsymbol{\sigma}$ is again obtained, uniform throughout the deforming domain.

This case can be directly extended to the case when the unit square is partitioned as an arbitrary number of uniform right angled triangles, such that, the resulting solution has the deformation gradient equal to $\textbf{F}$ on every deforming triangle except for the top right-hand side triangle, where the deformation gradient is $\widehat{\textbf{F}}$. Therefore, we conclude that there are infinitely many possible deformed states with non-homogeneous strain distribution giving the same homogeneous Cauchy stress throughout the elastic domain, provided that the Cauchy stress tensor given by (\ref{Eq:sigma}), or by (\ref{Eq:sigma:inc}) if the material is incompressible, can be expressed equivalently in terms of two different homogeneous left Cauchy-Green tensors $\textbf{B}=\textbf{F}\textbf{F}^{T}$ and $\widehat{\textbf{B}}=\widehat{\textbf{F}}\widehat{\textbf{F}}^{T}$, where $\textbf{F}$ and $\widehat{\textbf{F}}$ are rank-one connected.

%%%%%%%%%%%%%%%%%%%%%%%%%%%%%%%%%%
\subsubsection{Homogeneous deformation}

In order to obtain the homogeneous left Cauchy-Green tensor $\textbf{B}$ throughout the entire domain, the following boundary conditions can be prescribed:
\begin{itemize}
\item Homogeneous Dirichlet boundary conditions
\begin{eqnarray}
\textbf{u}(\textbf{X})=
\left[
\begin{array}{c}
a_{11}X_{1}+a_{12}X_{2}\\
a_{21}X_{1}+a_{22}X_{2}
\end{array}
\right] & \mbox{on} & (0,1)\times \{0,1\}\cup\{0,1\}\times(0,1),\label{Eq:dbc}
\end{eqnarray}
such that $u_{1}=u_{2}=0$ at $\textbf{X}=(0,0)$.
\end{itemize}

\begin{figure}[htbp]
\begin{center}
(a)\ \scalebox{0.42}{\includegraphics{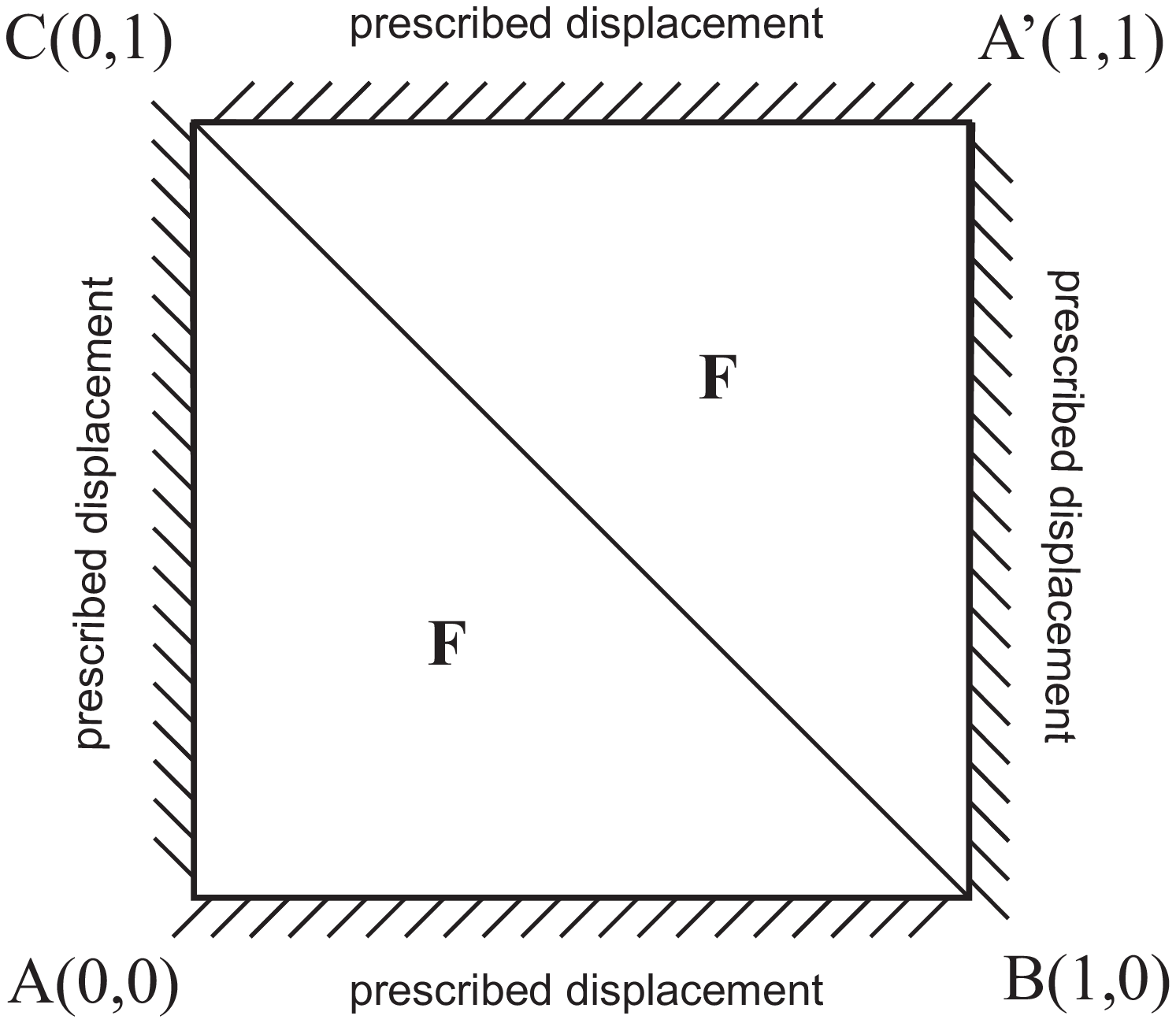}}\qquad
(b)\ \scalebox{0.42}{\includegraphics{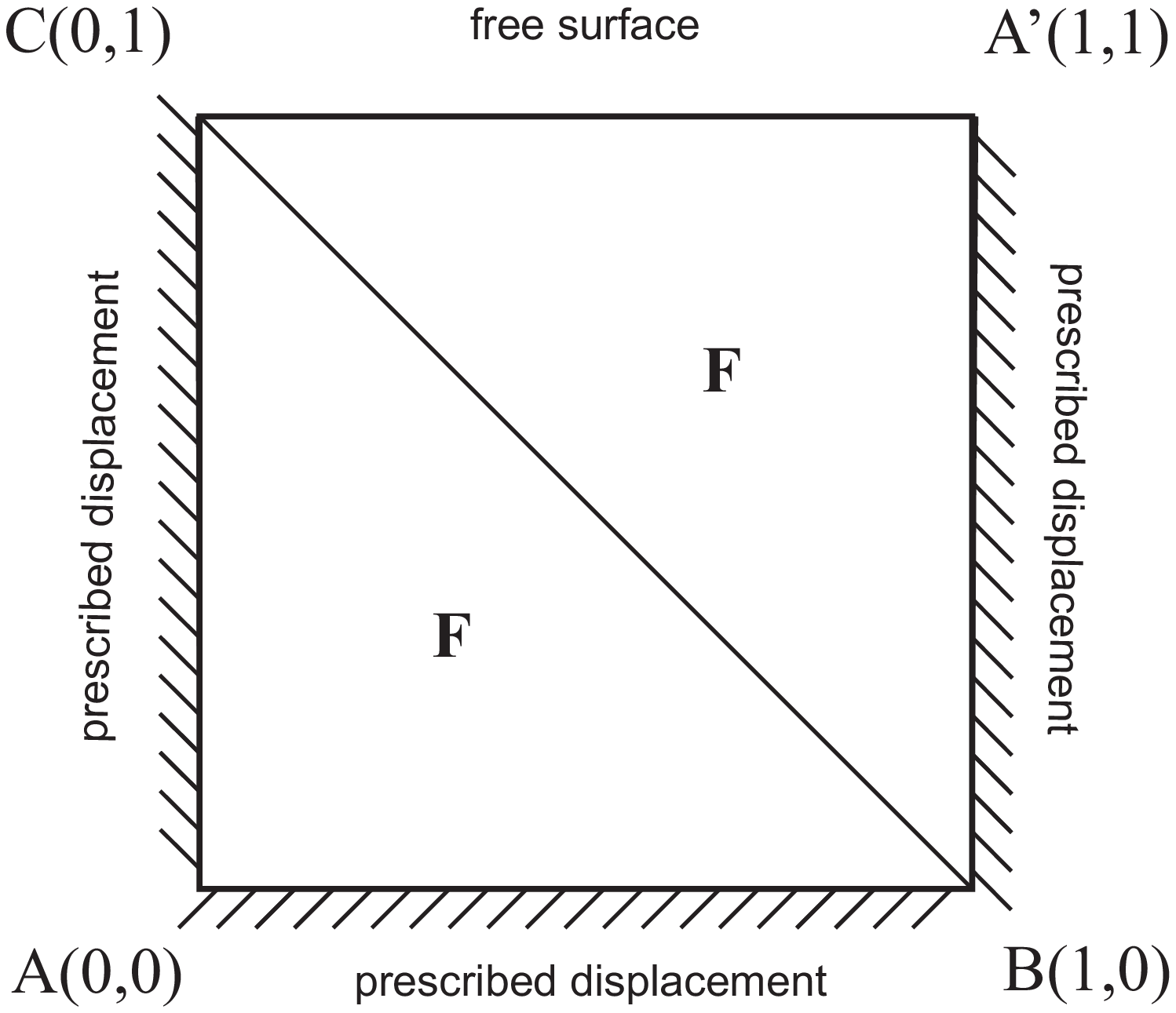}}
\caption{Elastic square with (a) uniform Dirichlet boundary conditions and (b) with one free side and Dirichlet boundary conditions on the remaining three sides, partitioned as two right-angled triangles, and the corresponding homogeneous deformation gradient indicated in each triangle.}\label{fig:usquare-triangles-dbc}
\end{center}
\end{figure}

Solving the equation (\ref{Eq:balance}) with the Dirichlet boundary conditions (\ref{Eq:dbc}) gives the left Cauchy-Green tensor $\textbf{B}$ and the first Piola-Kirchhoff stress tensors $\textbf{S}_{1}$ throughout the deforming domain, as indicated in Figure~\ref{fig:usquare-triangles-dbc} (a). Then the given Cauchy stress $\boldsymbol{\sigma}$ is produced throughout the deforming square.

%%%%%%%%%%%%%%%%
%%%%%%%%%%%%%%%%%

The same homogeneous solution is found when one side of the square is free and the Dirichlet boundary conditions (\ref{Eq:dbc}) are prescribed on the remaining three sides, as shown in Figure~\ref{fig:usquare-triangles-dbc} (b).

\begin{figure}[htbp]
\begin{center}
(a)\ \scalebox{0.42}{\includegraphics{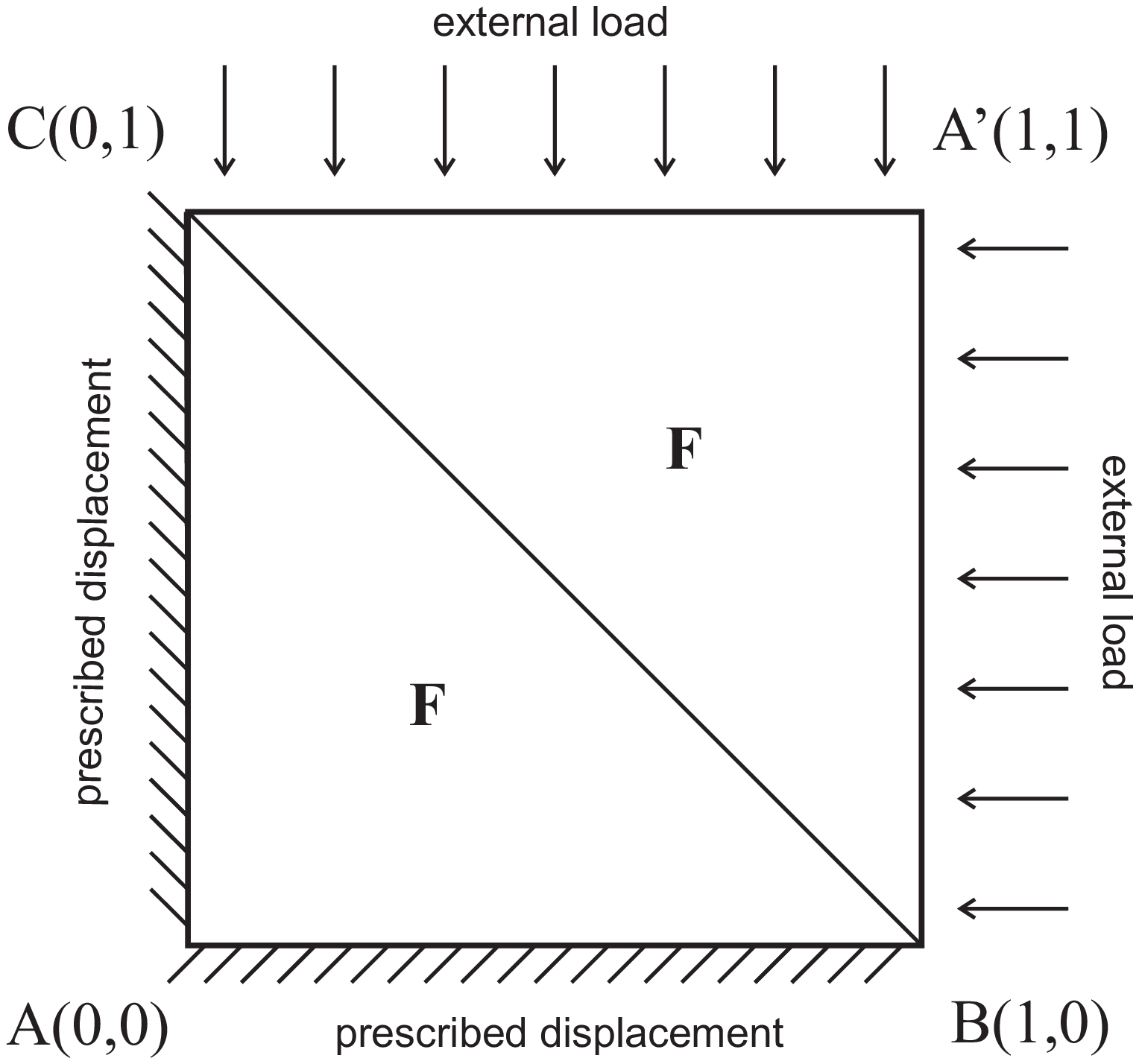}}\qquad
(b)\ \scalebox{0.42}{\includegraphics{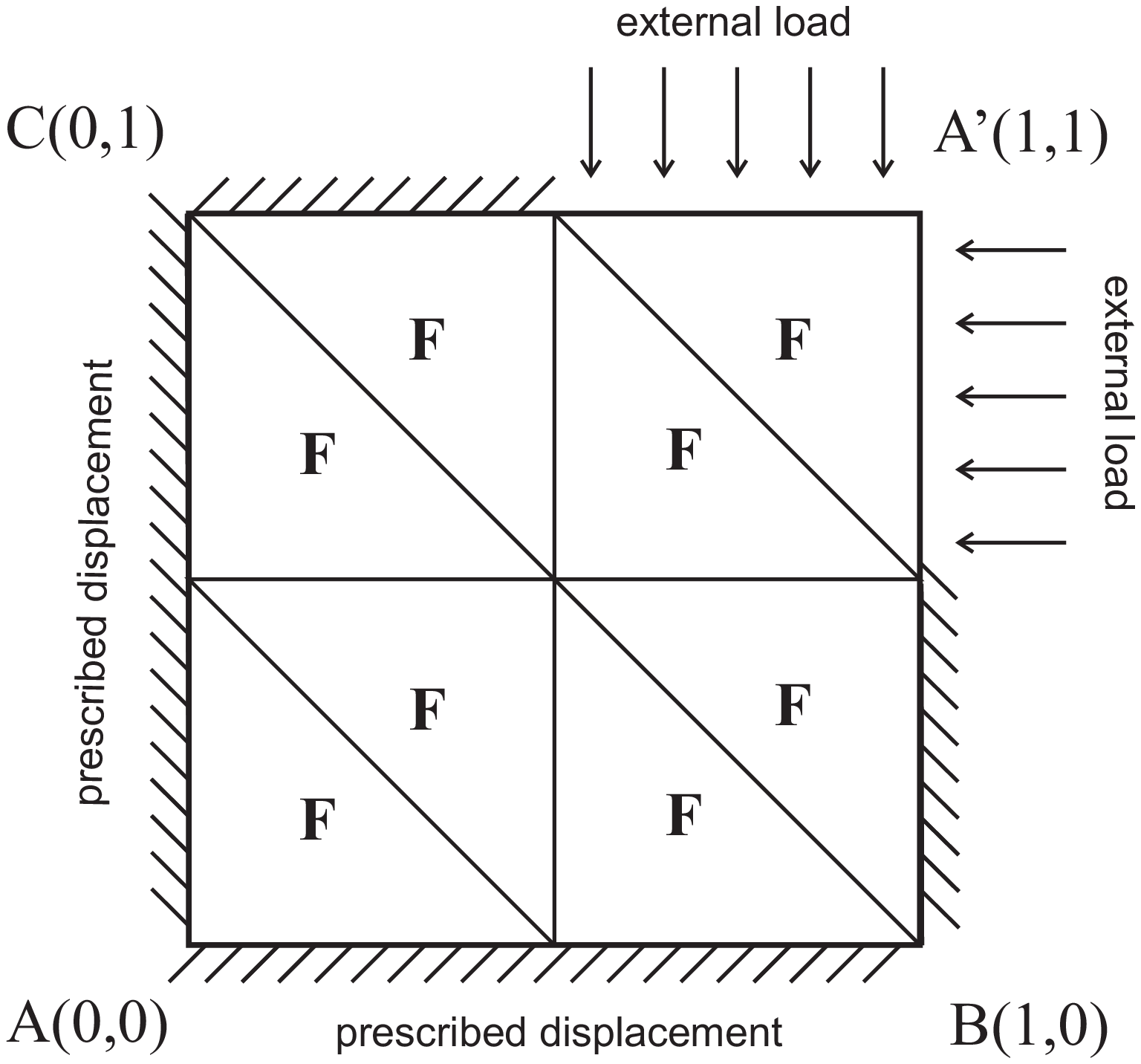}}
\caption{Elastic square with mixed boundary conditions, partitioned as two right-angled triangles, and the corresponding homogeneous deformation gradient indicated in each triangle.}\label{fig:usquare-triangles-mbc}
\end{center}
\end{figure}

\begin{itemize}
\item Alternatively, the above homogeneous solution can be obtained by imposing the following mixed boundary conditions, as indicated in Figure~\ref{fig:usquare-triangles-mbc} (a):
\begin{eqnarray}
\textbf{u}(\textbf{X})=
\left[
\begin{array}{c}
a_{11}X_{1}\\
a_{21}X_{1}
\end{array}
\right] & \mbox{on} & (0,1)\times \{0\},\label{Eq:mbc1:1}\\
\textbf{u}(\textbf{X})=
\left[
\begin{array}{c}
a_{12}X_{2}\\
a_{22}X_{2}
\end{array}
\right] & \mbox{on} & \{0\}\times(0,1),\label{Eq:mbc1:2}\\
\textbf{S}_{1}(\textbf{X})\left[
\begin{array}{r}
1\\
0
\end{array}
\right]=
\left[
\begin{array}{r}
S_{12}\\
S_{22}
\end{array}
\right]
& \mbox{on} & (0,1)\times \{1\},\label{Eq:mbc1:3}\\
\textbf{S}_{1}(\textbf{X})\left[
\begin{array}{r}
0\\
1
\end{array}
\right]=
\left[
\begin{array}{r}
S_{11}\\
S_{21}
\end{array}
\right] & \mbox{on} & \{1\}\times(0,1),\label{Eq:mbc1:4}
\end{eqnarray}
such that $u_{1}=u_{2}=0$ at $\textbf{X}=(0,0)$. In this case also, at a corner where one of the adjacent edges is subject to Dirichlet conditions and the other to Neumann conditions, the Dirichlet conditions take priority, and when both edges meeting at a corner are subject to Neumann conditions, these conditions are imposed simultaneously at the corner.
\end{itemize}

%%%%%%%%%%%%%%%%%
%%%%%%%%%%%%%%%

\begin{itemize}
\item Other mixed boundary conditions leading to the same homogeneous solution may also be given in the following form:
\begin{eqnarray}
\textbf{u}(\textbf{X})=
\left[
\begin{array}{c}
a_{11}X_{1}\\
a_{21}X_{1}
\end{array}
\right] & \mbox{on} & (0,1)\times \{0\},\label{Eq:mbc2:1}\\
\textbf{u}(\textbf{X})=
\left[
\begin{array}{c}
a_{12}X_{2}\\
a_{22}X_{2}
\end{array}
\right] & \mbox{on} & \{0\}\times(0,1),\label{Eq:mbc2:2}\\
\textbf{u}(\textbf{X})=
\left[
\begin{array}{c}
a_{11}X_{1}+a_{12}\\
a_{21}X_{1}+a_{22}
\end{array}
\right] & \mbox{on} &,\label{Eq:mbc2:3}\\
\textbf{u}(\textbf{X})=
\left[
\begin{array}{c}
a_{11}+a_{12}X_{2}\\
a_{21}+a_{22}X_{2}
\end{array}
\right] & \mbox{on} & \{1\}\times(0,1/2),\label{Eq:mbc2:4}\\
\textbf{S}_{1}(\textbf{X})\left[
\begin{array}{r}
1\\
0
\end{array}
\right]=
\left[
\begin{array}{r}
S_{12}\\
S_{22}
\end{array}
\right]
& \mbox{on} & (1/2,1)\times \{1\},\label{Eq:mbc2:5}\\
\textbf{S}_{1}(\textbf{X})\left[
\begin{array}{r}
0\\
1
\end{array}
\right]=
\left[
\begin{array}{r}
S_{11}\\
S_{21}
\end{array}
\right] & \mbox{on} & \{1\}\times(1/2,1),\label{Eq:mbc2:6}
\end{eqnarray}
such that $u_{1}=u_{2}=0$ at $\textbf{X}=(0,0)$. At a boundary point where one of the adjacent edges is subject to Dirichlet conditions and the other to Neumann conditions, the Dirichlet conditions take priority, and when both edges meeting at a point are subject to Neumann conditions, these conditions are imposed simultaneously at that point.
\end{itemize}

This case is illustrated graphically in Figure~\ref{fig:usquare-triangles-mbc} (b), and can be extended directly to the case when the unit square is partitioned as an arbitrary number of uniform right angled triangles.

%%%%%%%%%%%%%%%%%%%%%%%%%%%%%%%%%%%%%%%%%%%%%%%%%%%%%%%%%%%%
%%%%%%%%%%%%%%%%%%%%   NEW SECTION  %%%%%%%%%%%%%%%%%%%%%%%%
%%%%%%%%%%%%%%%%%%%%%%%%%%%%%%%%%%%%%%%%%%%%%%%%%%%%%%%%%%%%
\section{Deriving suitable deformations}\label{sec:example}

Given the strain energy density function for a homogeneous isotropic hyperelastic material, suitable elastic deformations can be found, such that the corresponding Cauchy stress tensor can be expressed equivalently in terms of two different homogeneous left Cauchy-Green tensors $\textbf{B}=\textbf{F}\textbf{F}^{T}$ and $\widehat{\textbf{B}}=\widehat{\textbf{F}}\widehat{\textbf{F}}^{T}$, where $\textbf{F}$ and $\widehat{\textbf{F}}$ take the form (\ref{Eq:F}) and $(\ref{Eq:Fhat})$, respectively, and are rank-one connected.

For unconstrained materials, writing the components of the Cauchy stress described by (\ref{Eq:sigma}) in the two equivalent forms leads to the following three simultaneous equations:
\begin{eqnarray}
\beta_{0}(\textbf{B})+\beta_{1}(\textbf{B})B_{11}+\beta_{-1}(\textbf{B})B_{22}/I_{3}(\textbf{B})&=&\beta_{0}(\widehat{\textbf{B}})+\beta_{1}(\widehat{\textbf{B}})\widehat{B}_{11}+\beta_{-1}(\widehat{\textbf{B}})\widehat{B}_{22}/I_{3}(\widehat{\textbf{B}}),\label{Eq:model:stress1}\\
\beta_{0}(\textbf{B})+\beta_{1}(\textbf{B})B_{22}+\beta_{-1}(\textbf{B})B_{11}/I_{3}(\textbf{B})&=&\beta_{0}(\widehat{\textbf{B}})+\beta_{1}(\widehat{\textbf{B}})\widehat{B}_{22}+\beta_{-1}(\widehat{\textbf{B}})\widehat{B}_{11}/I_{3}(\widehat{\textbf{B}})\label{Eq:model:stress2},\\
\left[\beta_{1}(\textbf{B})-\beta_{-1}(\textbf{B})/I_{3}(\textbf{B})\right]B_{12}&=&\left[\beta_{1}(\widehat{\textbf{B}})-\beta_{-1}(\widehat{\textbf{B}})/I_{3}(\widehat{\textbf{B}})\right]\widehat{B}_{12}\label{Eq:model:stress12},
\end{eqnarray}
where:
\begin{eqnarray}
B_{11}&=&F_{11}^2+F_{12}^2,\label{Eq:B11}\\
B_{12}&=&F_{11}F_{21}+F_{12}F_{22},\label{Eq:B12}\\
B_{22}&=&F_{21}^2+F_{22}^2,\label{Eq:B22}
\end{eqnarray}
and
\begin{eqnarray}
\widehat{B}_{11}&=&\widehat{F}_{11}^2+\widehat{F}_{12}^2,\label{Eq:widehatB11}\\
\widehat{B}_{12}&=&\widehat{F}_{11}\widehat{F}_{21}+\widehat{F}_{12}\widehat{F}_{22},\label{Eq:widehatB12}\\
\widehat{B}_{22}&=&\widehat{F}_{21}^2+\widehat{F}_{22}^2.\label{Eq:widehatB22}
\end{eqnarray}

The rank-one connectivity condition means
\begin{eqnarray}
\left(F_{11}-\widehat{F}_{11}\right)\left(F_{22}-\widehat{F}_{22}\right)&=&\left(F_{12}-\widehat{F}_{12}\right)\left(F_{21}-\widehat{F}_{21}\right).\label{Eq:model:rankone}
\end{eqnarray}

From the four nonlinear equations (\ref{Eq:model:stress1})-(\ref{Eq:model:stress12}) and (\ref{Eq:model:rankone}), the components $\{\widehat{F}_{11}, \widehat{F}_{12},\widehat{F}_{21},\widehat{F}_{22}\}$ of the deformation gradient $\widehat{\textbf{F}}$ can be determined, at least in principle, in terms of the components $\{F_{11},F_{12},F_{21},F_{22}\}$ of the deformation gradient $\textbf{F}$.

For incompressible materials, the components of the Cauchy stress described by (\ref{Eq:sigma:inc}) expressed in the two equivalent forms leads to the following three simultaneous equations:
\begin{eqnarray}
p_{0}+\beta_{1}(\textbf{B})B_{11}+\beta_{-1}(\textbf{B})B_{22}&=&\widehat{p}_{0}+\beta_{1}(\widehat{\textbf{B}})\widehat{B}_{11}+\beta_{-1}(\widehat{\textbf{B}})\widehat{B}_{22},\label{Eq:model:stress1:inc}\\
p_{0}+\beta_{1}(\textbf{B})B_{22}+\beta_{-1}(\textbf{B})B_{11}&=&\widehat{p}_{0}+\beta_{1}(\widehat{\textbf{B}})\widehat{B}_{22}+\beta_{-1}(\widehat{\textbf{B}})\widehat{B}_{11}\label{Eq:model:stress2:inc},\\
\left[\beta_{1}(\textbf{B})-\beta_{-1}(\textbf{B})\right]B_{12}&=&\left[\beta_{1}(\widehat{\textbf{B}})-\beta_{-1}(\widehat{\textbf{B}})\right]\widehat{B}_{12}\label{Eq:model:stress12:inc},
\end{eqnarray}
where the components of the left Cauchy-Green tensors $\textbf{B}$ and $\widehat{\textbf{B}}$ are given by (\ref{Eq:B11})-(\ref{Eq:B22}) and (\ref{Eq:widehatB11})-(\ref{Eq:widehatB22}), respectively, and $p_{0}$ and $\widehat{p}_{0}$ are the associated hydrostatic pressures. 

In this case, in addition to the condition (\ref{Eq:model:rankone}), the following incompressibility constraints must be satisfied:
\begin{eqnarray}
F_{11}F_{22}-F_{12}F_{21}&=&0,\label{Eq:model:inc}\\
\widehat{F}_{11}\widehat{F}_{22}-\widehat{F}_{12}\widehat{F}_{21}&=&0.\label{Eq:model:wideinc}
\end{eqnarray}

From the equations (\ref{Eq:model:stress1:inc})-(\ref{Eq:model:stress12:inc}), (\ref{Eq:model:rankone}) and (\ref{Eq:model:inc})-(\ref{Eq:model:wideinc}), the components $\{\widehat{F}_{11}, \widehat{F}_{12},\widehat{F}_{21},\widehat{F}_{22}\}$ of the deformation gradient $\widehat{\textbf{F}}$ and the hydrostatic pressure $\widehat{p}_{0}$ can be determined  in terms of the components $\{F_{11},F_{12},F_{21},F_{22}\}$ of the deformation gradient $\textbf{F}$ and the hydrostatic pressure $p_{0}$.

\begin{example}
We offer a simple example of two homogeneous deformations leading to the same Cauchy stress in a given unconstrained homogeneous isotropic hyperelastic material characterised by the following strain energy density function
\begin{equation}\label{Eq:model:W}
\begin{split}
W&=\frac{\mu}{2}\left(I_{3}^{-1/3}I_{1}-3\right)+\frac{\tilde{\mu}}{4}\left(I_{1}-3\right)^2+\frac{\kappa}{2}\left(I_{3}^{1/2}-1\right)^2\\
&=\frac{\mu}{2}\left[\left\|\frac{\textbf{F}}{\left(\det\textbf{F}\right)^{1/3}}\right\|^2-3\right]+\frac{\tilde{\mu}}{4}\left(\|\textbf{F}\|^2-3\right)^2+\frac{\kappa}{2}\left(\det\textbf{F}-1\right)^2,
\end{split}
\end{equation}
where $\mu>0$ is the infinitesimal shear modulus, $\kappa>0$ is the infinitesimal bulk modulus, $\tilde{\mu}$ is an additional positive constant independent of the deformation, and $\|\cdot\|$ is the Frobenius norm. This energy function is not rank-one convex due to the presence of the $\tilde{\mu}$-term. Nevertheless, it is $LH$-elliptic in a neighbourhood of the identity \cite{Hartmann:2003:HN}.

For the material model (\ref{Eq:model:W}), differentiating with respect to the strain invariants gives:
\begin{eqnarray*}
\frac{\partial W}{\partial I_{1}}&=&\frac{\mu}{2}I_{3}^{-1/3}+\frac{\tilde{\mu}}{2}\left(I_{1}-3\right),\\
\frac{\partial W}{\partial I_{2}}&=&0,\\
\frac{\partial W}{\partial I_{3}}&=&-\frac{\mu}{6}I_{1}I_{3}^{-4/3}+\frac{\kappa}{2}I_{3}^{-1/2}\left(I_{3}^{1/2}-1\right),
\end{eqnarray*}
and the coefficients (\ref{Eq:betas}) take the form:
\begin{equation}\label{Eq:betas:example}
\beta_{0}=-\frac{\mu}{3}I_{1}I_{3}^{-5/6}+\kappa\left(I_{3}^{1/2}-1\right),\qquad
\beta_{1}=\mu I_{3}^{-5/6}+\tilde{\mu}I_{3}^{-1/2}\left(I_{1}-3\right),\qquad
\beta_{-1}=0.
\end{equation}

We consider two homogeneous deformations with the following deformation gradients
\begin{equation}\label{Eq:FhatF:example}
\textbf{F}=
\left[
\begin{array}{ccc}
k & s & 0\\
0 & 1 & 0\\
0 & 0 & 1
\end{array}
\right],
\qquad
\widehat{\textbf{F}}=
\left[
\begin{array}{ccc}
k & -s & 0\\
0 & 1 & 0\\
0 & 0 & 1
\end{array}
\right],
\end{equation}
where $k$ and $s$ are positive constants, hence the rank-one connectivity condition (\ref{Eq:rank1}) is satisfied. 

The corresponding left Cauchy-Green tensors are, respectively,
\begin{equation}\label{Eq:BhatB:example}
\textbf{B}=\textbf{F}\textbf{F}^{T}=
\left[
\begin{array}{ccc}
k^2+s^2 & s & 0\\
s & 1 & 0\\
0 & 0 & 1
\end{array}
\right],
\qquad
\widehat{\textbf{B}}=\widehat{\textbf{F}}\widehat{\textbf{F}}^{T}=
\left[
\begin{array}{ccc}
k^2+s^2 & -s & 0\\
-s & 1 & 0\\
0 & 0 & 1
\end{array}
\right],
\end{equation}
and, by (\ref{Eq:sigma}) and (\ref{Eq:betas:example}), the associated Cauchy stresses take the form
\begin{equation}\label{Eq:sigmas:example}
\boldsymbol{\sigma}(\textbf{B})=\beta_{0}\ \textbf{I}+\beta_{1}\ \textbf{B},
\qquad
\boldsymbol{\sigma}(\widehat{\textbf{B}})=\beta_{0}\ \textbf{I}+\beta_{1}\ \widehat{\textbf{B}}.
\end{equation}

For these deformations, $\det\textbf{F}=\det\widehat{\textbf{F}}=k>0$, and if $s\neq0$, then $\textbf{F}\neq\widehat{\textbf{F}}$ and $\textbf{B}\neq\widehat{\textbf{B}}$. A graphical illustration of such deformations is shown in Figure~\ref{fig:deform-example}.

%%%%%%%%%%%%%%%%%
\begin{figure}[htbp]
\begin{center}
(a) \scalebox{0.4}{\includegraphics{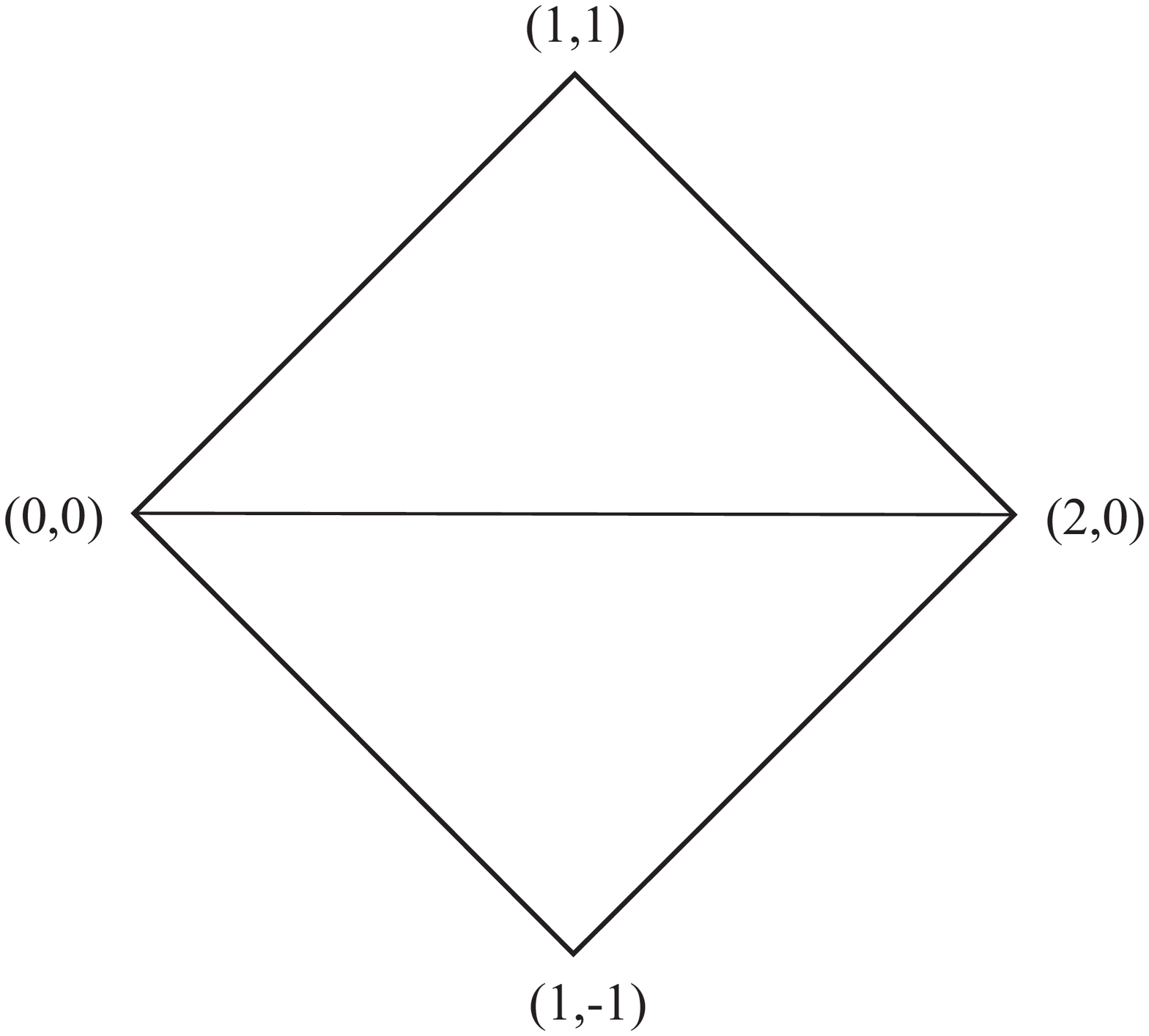}}\\
(b) \scalebox{0.4}{\includegraphics{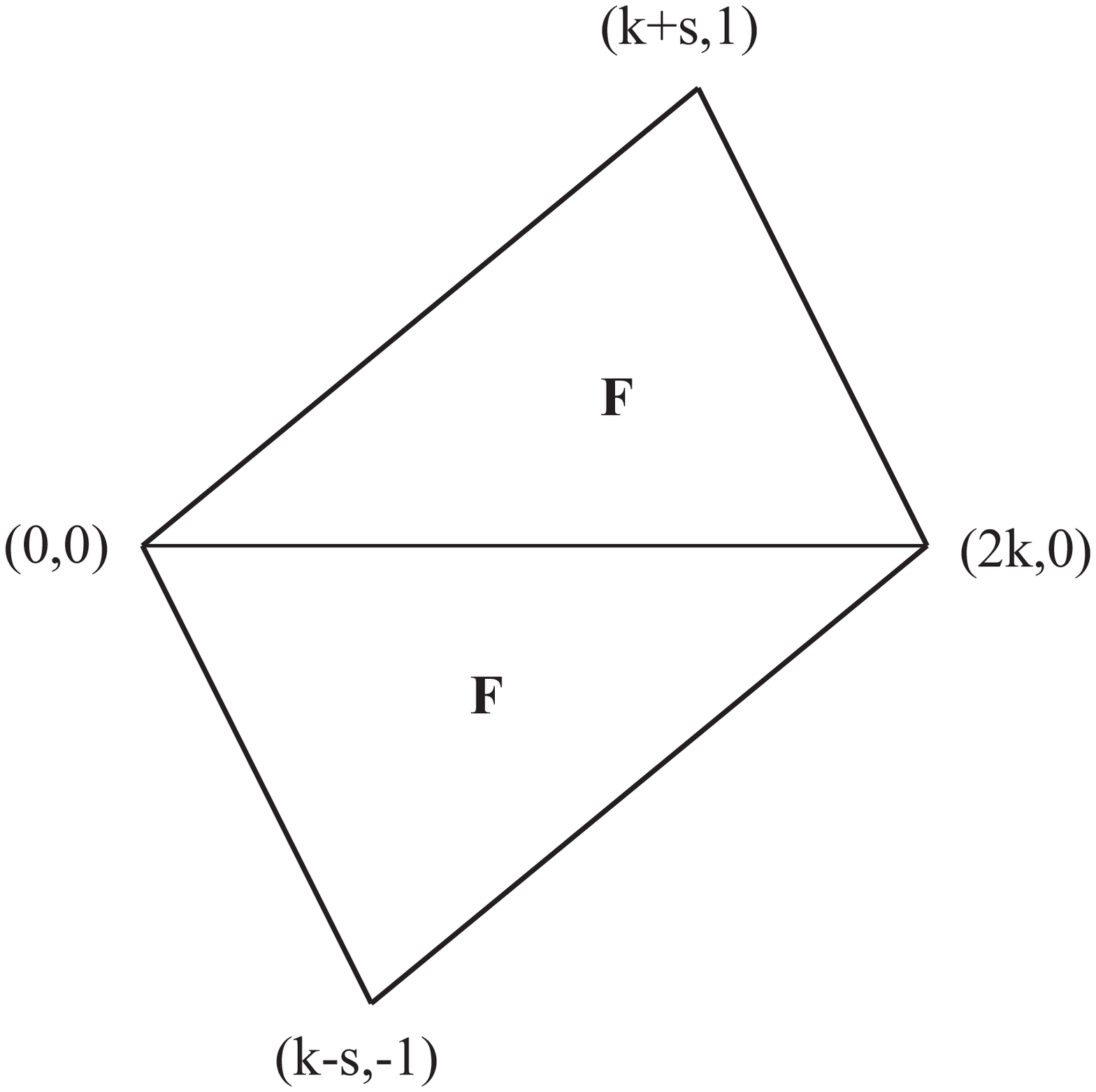}}
(c) \scalebox{0.4}{\includegraphics{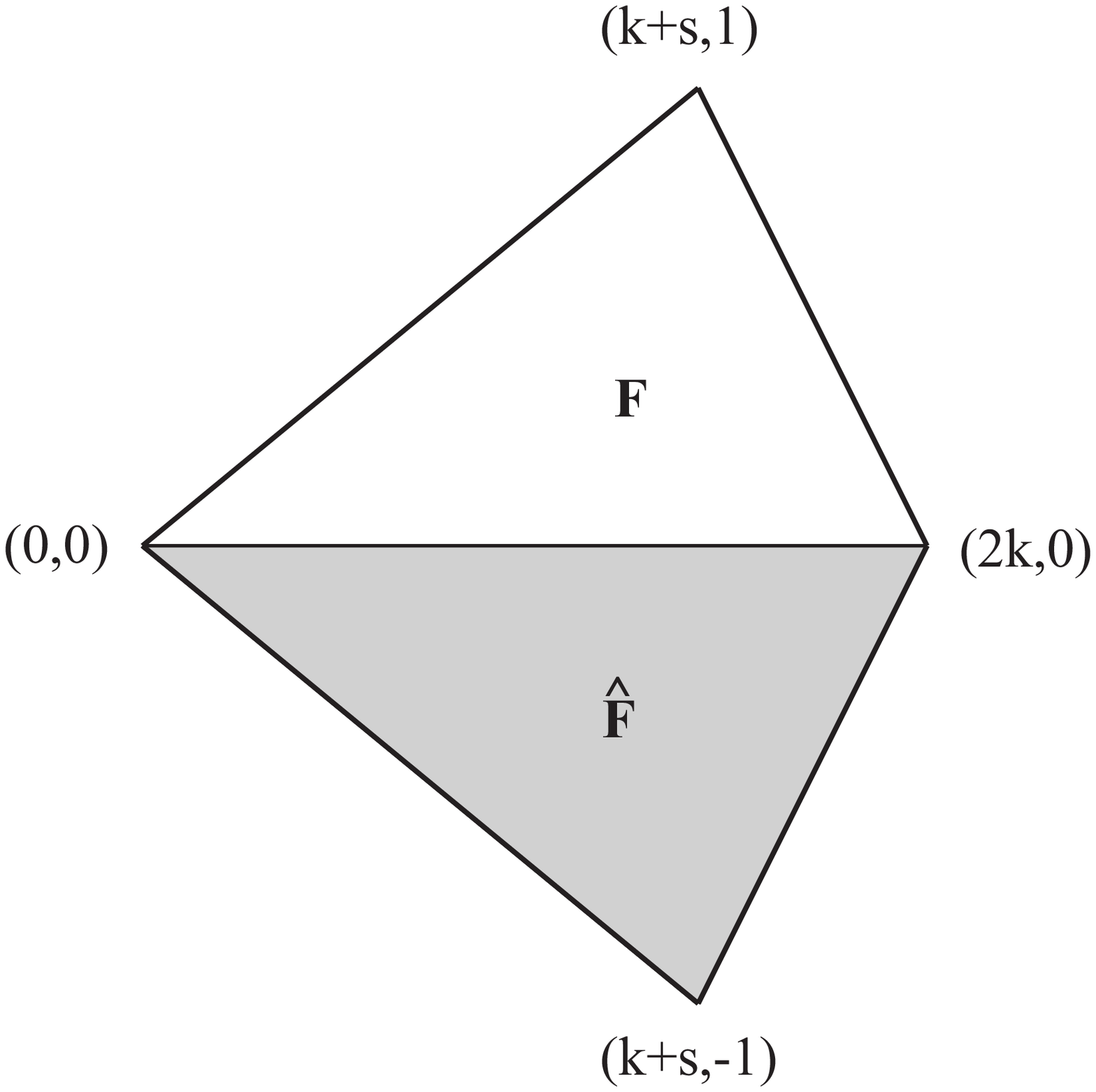}}
\caption{Elastic square partitioned into two right-angled triangles, showing: (a) the reference configuration, (b) the deformed state with the deformation gradient equal to $\textbf{F}$ on each triangle, (c) the deformed state with the deformation gradient equal to $\textbf{F}$ on one triangle and $\widehat{\textbf{F}}$ on the other.}\label{fig:deform-example}
\end{center}
\end{figure}
%%%%%%%%%%%%%%%

Since $I_{1}=k^2+s^2+2$ and $I_{3}=k^2$ are invariants for both $\textbf{B}$ and $\widehat{\textbf{B}}$, it follows that, in (\ref{Eq:sigmas:example}),
\[
\beta_{0}=-\frac{\mu}{3}k^{-5/3}\left(k^2+s^2+2\right)+\kappa\left(k-1\right),\qquad
\beta_{1}=\mu k^{-5/3}+\tilde{\mu}k^{-1}\left(k^2+s^2-1\right).
\]

It can be verified that, if $\mu/\left(3\tilde{\mu}\right)<4^{-4/3}$ and $0<s<\sqrt{1-4\left[\mu/\left(3\tilde{\mu}\right)\right]^{3/4}}$, then there exists $k_{0}\in(0,1)$, such that, for $k=k_{0}$,
\[
\beta_{0}=-\frac{\mu}{3}k_{0}^{-5/3}\left(k_{0}^2+s^2+2\right)-\kappa\left(1-k_{0}\right)<0,\qquad
\beta_{1}=0,
\]
and all the equations (\ref{Eq:model:stress1})-(\ref{Eq:model:stress12}) and (\ref{Eq:model:rankone}) are satisfied, with the common Cauchy stress tensor produced by these deformations taking the form
\begin{eqnarray*}
\boldsymbol{\sigma}(\textbf{B})=\boldsymbol{\sigma}(\widehat{\textbf{B}})=\beta_{0}\ \textbf{I}.
\end{eqnarray*}
\end{example}

Note that, our example does not violate the uniqueness result from linear elasticity even if $s>0$ is small. If $s\to0$ and $k\to1$, corresponding to the linear elastic limit in (\ref{Eq:BhatB:example}), then $s^2+k^2-1$ is arbitrarily small, and $\beta_{1}=\mu k^{-5/3}+\tilde{\mu}k^{-1}\left(k^2+s^2-1\right)\to\mu\neq0$. Hence if $\beta_{1}=0$ and $s$ is close to zero, then $k$ cannot be close to one, and therefore the two different deformation gradients (\ref{Eq:FhatF:example}), which are rank-one connected, do not correspond to infinitesimal deformations.

Furthermore, if different $k_{1},k_{2}\in(0,1)$ exist, such that $\beta_{1}=0$ with the same $s>0$, then two different deformation gradients
\[
\textbf{F}=
\left[
\begin{array}{ccc}
k_{1} & s & 0\\
0 & 1 & 0\\
0 & 0 & 1
\end{array}
\right]
\qquad\mbox{and}\qquad
\widehat{\textbf{F}}=
\left[
\begin{array}{ccc}
k_{1} & -s & 0\\
0 & 1 & 0\\
0 & 0 & 1
\end{array}
\right],
\]
satisfy (\ref{Eq:rank1}) and produce the same Cauchy stress
\[
\boldsymbol{\sigma}=\beta_{0}\ \textbf{I}=\left[-\frac{\mu}{3}k_{1}^{-5/3}\left(k_{1}^2+s^2+2\right)-\kappa\left(1-k_{1}\right)\right]\textbf{I}, 
\]
and similarly,
\[
\textbf{F}=
\left[
\begin{array}{ccc}
k_{2} & s & 0\\
0 & 1 & 0\\
0 & 0 & 1
\end{array}
\right]
\qquad\mbox{and}\qquad
\widehat{\textbf{F}}=
\left[
\begin{array}{ccc}
k_{2} & -s & 0\\
0 & 1 & 0\\
0 & 0 & 1
\end{array}
\right],
\]
are rank-one connected and produce the Cauchy stress
\[
\boldsymbol{\sigma}=\beta_{0}\ \textbf{I}=\left[-\frac{\mu}{3}k_{2}^{-5/3}\left(k_{2}^2+s^2+2\right)-\kappa\left(1-k_{2}\right)\right]\textbf{I}.
\]

Looking back at our example (\ref{Eq:model:W}), we have constructed an elastic strain energy which is not rank-one convex and which allows for inhomogeneous deformations leading to a homogeneous Cauchy stress. This leads to the following question: is it possible to find a rank-one convex elastic energy, such that the Cauchy stress $\boldsymbol{\sigma}$ is not injective and there exists a homogeneous state with deformation gradient $\textbf{F}$, such that $\sigma(\textbf{F})=\sigma(\textbf{F}+\textbf{a}\otimes\textbf{n})$, with $\textbf{a}$ and $\textbf{n}$ as given in (\ref{Eq:rank1an}). The answer to this question, however, is negative, and we show this in \cite{Neff:2016:NM}.

%%%%%%%%%%%%%%%%%%%%%%%%%%%%%%%%%%%%%%%%%%%%%%%%%%%%%%%%%%%%
%%%%%%%%%%%%%%%%%%%%   NEW SECTION  %%%%%%%%%%%%%%%%%%%%%%%%
%%%%%%%%%%%%%%%%%%%%%%%%%%%%%%%%%%%%%%%%%%%%%%%%%%%%%%%%%%%%
\section{Conclusion}

We established here that, in isotropic finite elasticity, unlike in the linear elastic theory, homogeneous Cauchy stress does not imply homogeneous strain. To demonstrate this, we first identified such situations with compatible, continuous deformations on a specific geometry. Then we provided an example of an isotropic strain energy function, such that, if a material is described by this function and occupies a domain similar to those analysed, then the expressions of the homogeneous Cauchy stress and the corresponding non-homogeneous strains could be written explicitly. We derived our example from a non rank-one convex elastic energy.

%%%%%%%%%%%%%%%%%%%%%%%%%%%%%%%%%%%%%%%%%%%%%%%%%%%%%%%%%%%%
%%%%%%%%%%%%%%%%%%%%   NEW SECTION  %%%%%%%%%%%%%%%%%%%%%%%%
%%%%%%%%%%%%%%%%%%%%%%%%%%%%%%%%%%%%%%%%%%%%%%%%%%%%%%%%%%%%
\section*{Acknowledgements}
The support for L. Angela Mihai by the Engineering and Physical Sciences Research Council of Great Britain under research grant EP/M011992/1 is gratefully acknowledged. 

%%%%%%%%%%%%%%%%%%%%%%%%%%%%%%%%%%%%%%%%%%%%%%%%%%%%%%%%%%%%
%%%%%%%%%%%%%%%%%%%%  NEW SECTION   %%%%%%%%%%%%%%%%%%%%%%%%
%%%%%%%%%%%%%%%%%%%%%%%%%%%%%%%%%%%%%%%%%%%%%%%%%%%%%%%%%%%%

\end{document}